%% file: main.tex
\documentclass[10pt,journal,compsoc]{IEEEtran}

\usepackage[english]{babel}
\usepackage[utf8]{inputenc}
\usepackage[nocompress]{cite}
\usepackage[pdftex]{graphicx}
\usepackage{siunitx}
\usepackage{amsmath}
\usepackage{amssymb}
\usepackage{wasysym}
\usepackage{array}
\usepackage{tikz}
\usepackage[caption=false,font=footnotesize,labelfont=sf,textfont=sf]{subfig}
\usepackage{float}
\usepackage{adjustbox}
\usepackage[acronym]{glossaries}
\usepackage[hidelinks]{hyperref}
\usepackage[capitalise, noabbrev]{cleveref}
\usepackage{nicefrac}
\usepackage{listings}
\usepackage{lipsum}
\usepackage{calc}
\usepackage[font=footnotesize,labelfont=sf,textfont=sf]{caption}
\usepackage{booktabs}
\usepackage{ragged2e}
\usepackage{multirow}
\usepackage[symbol]{footmisc}
\usepackage{mdframed}
\usepackage{xcolor}
\usepackage{textgreek}

\input{acr}

\graphicspath{{fig/}}
\DeclareSIUnit\GE{GE}
\DeclareSIUnit\kGE{\kilo\GE}
\DeclareSIUnit\MGE{\mega\GE}
\DeclareSIUnit\flop{flop}
\DeclareSIUnit\flops{\flop\per\second}
\DeclareSIUnit\Gflops{\giga\flops}
\DeclareSIUnit\Tflops{\tera\flops}
\newfloat{listing}{htbp}{lst}
\floatname{listing}{Listing}
\newcounter{sublisting}
\newcounter{sublisting@save}
\crefname{sublisting}{listing}{listings}
\Crefname{sublisting}{Listing}{Listings}

\newcommand{\riscv}{RISC-V}
\newcommand{\x}{$\times$}

\newcommand{\topar}[1]{\textit{#1}:}

\def\showieeenotice{}

\begin{document}


\title{Sparse Stream Semantic Registers:\\A Lightweight ISA Extension Accelerating\\General Sparse Linear Algebra}

\author{%
  Paul~Scheffler,~\IEEEmembership{Student Member,~IEEE,}
  Florian~Zaruba,
  Fabian~Schuiki,\\
  Torsten~Hoefler,~\IEEEmembership{Fellow,~IEEE,}
  Luca~Benini,~\IEEEmembership{Fellow,~IEEE}
  \IEEEcompsocitemizethanks{%
    \IEEEcompsocthanksitem P. Scheffler and L. Benini are with the Integrated Systems Laboratory (IIS), ETH Zurich. E-mail: \{paulsc,lbenini\}@iis.ee.ethz.ch
    \IEEEcompsocthanksitem F. Zaruba is with Axelera AI. Email: florian.zaruba@axelera.ai
    \IEEEcompsocthanksitem F. Schuiki is with SiFive. Email: fabian@schuiki.ch
    \IEEEcompsocthanksitem T. Hoefler  is with the Scalable Parallel Computing Laboratory (SPCL), ETH Zurich. Email: htor@inf.ethz.ch
    \IEEEcompsocthanksitem L. Benini is also with the Department of Electrical, Electronic, and Information Engineering (DEI), University of Bologna.
\ifx\showieeenotice\undefined
\else
    \IEEEcompsocthanksitem{\copyright~2023 IEEE.  Personal use of this material is permitted.  Permission from IEEE must be obtained for all other uses, in any current or future media, including reprinting/republishing this material for advertising or promotional purposes, creating new collective works, for resale or redistribution to servers or lists, or reuse of any copyrighted component of this work in other works.}%
    \IEEEcompsocthanksitem{IEEE Publication DOI: 10.1109/TPDS.2023.3322029}
\fi
  }%
}%

\ifx\showrebuttal\undefined
    \newcommand{\revone}[1]{{#1}}%
    \newcommand{\revonemod}[1]{{#1}}%
    \newenvironment{revonefig}[1][htbp]{\figure[#1]}{\endfigure}
    \newenvironment{revonefig*}[1][htbp]{\csname figure*\endcsname [#1]}{\csname endfigure*\endcsname}
\else
    \newcommand{\revone}[1]{{\textcolor{blue}{#1}}}
    \newcommand{\revonemod}[1]{{\textcolor{blue}{#1}}}
    \newenvironment{revonefig}[1][htbp]{\figure[#1]\mdframed[backgroundcolor=blue!15, topline=false, rightline=false, leftline=false, bottomline=false, innertopmargin=0, innerbottommargin=0, innerrightmargin=0,innerleftmargin=0]}{\endmdframed\endfigure}
    \newenvironment{revonefig*}[1][htbp]{\csname figure*\endcsname [#1]\mdframed[backgroundcolor=blue!15, topline=false, rightline=false, leftline=false, bottomline=false, innertopmargin=0, innerbottommargin=0, innerrightmargin=0,innerleftmargin=0]}{\endmdframed\csname endfigure*\endcsname}
\fi

\IEEEtitleabstractindextext{%
  \justify
  \begin{abstract}
Sparse linear algebra is crucial in many application domains, but challenging to handle efficiently in both software and hardware, with one- and two-sided operand sparsity handled with distinct approaches.
In this work, we enhance an existing memory-streaming RISC-V ISA extension to accelerate both one- and two-sided operand sparsity on widespread sparse tensor formats like compressed sparse row (CSR) and compressed sparse fiber (CSF) by accelerating the underlying operations of streaming indirection, intersection, and union. 
Our extensions enable single-core speedups over an optimized RISC-V baseline of up to 7.0x, 7.7x, and 9.8x on sparse-dense multiply, sparse-sparse multiply, and sparse-sparse addition, respectively, and peak FPU utilizations of up to 80\% on sparse-dense problems. 
On an eight-core cluster, sparse-dense and sparse-sparse matrix-vector multiply using real-world matrices are up to \revone{4.9x} and \revonemod{5.9x} faster and up to \revonemod{2.9x} and \revonemod{3.0x} more energy efficient.
We explore further applications for our extensions, such as stencil codes and graph pattern matching. Compared to recent CPU, GPU, and accelerator approaches, our extensions enable higher flexibility on data representation, degree of sparsity, and dataflow at a minimal hardware footprint, adding only 1.8\% in area to a compute cluster. A cluster with our extensions running CSR matrix-vector multiplication achieves \revone{9.9x}~and \revonemod{1.7x}~higher peak floating-point utilizations than recent \revone{highly optimized sparse data structures and libraries for} CPU and GPU, respectively, \revone{even when accounting for off-chip main memory (HBM) and on-chip interconnect latency and bandwidth effects.}
  \end{abstract}
}

\maketitle
\IEEEdisplaynontitleabstractindextext
\IEEEpeerreviewmaketitle


\IEEEraisesectionheading{\section{Introduction}\label{sec:introduction}}

\IEEEPARstart{S}{parse} \gls{la} poses a formidable challenge for contemporary processor architectures optimized for highly regular and vectorizable workloads. 
In sparse tensors, a significant portion of entries is zero. To reduce their memory footprint and eliminate redundant computation, they are usually compressed by storing only the values and positions of nonzero elements.
However, this results in \revone{nested indirect data structures that cannot be randomly accessed, irregular memory access patterns, and significant control overheads}.
Even the most fundamental sparse tensor operations, \emph{addition} and \emph{multiplication}, are inherently irregular. Worse, their optimal execution strategy depends on operand sparsity: while one sparse operand can be handled through indirect memory accesses, two sparse operands require comparing nonzero positions in both tensors.

Consider the multiplication of a sparse matrix \verb|A|, stored in the widespread \gls{csr} format, with a dense vector \verb|b|, which can be implemented as shown in \Cref{lst:intro_a}.
Here, \verb|A_vals| and \verb|A_idcs| store the nonzero values and their column indices, and \verb|A_ptrs| delimits the rows of \verb|A|. For each element of the result vector \verb|c|, we perform a \emph{sparse-dense} dot product \verb|sV×dV|\footnote{We henceforth use the prefixes \texttt{s} and \texttt{d} to denote \emph{sparse} and \emph{dense} tensors, respectively. Analogously, \texttt{M} denotes a \emph{matrix} and \texttt{V} a \emph{vector}.} as shown.
Compiling this \verb|sV×dV| for the \riscv{} instruction set yields a loop of nine instructions, of which only one, a fused \gls{mac}, performs necessary compute \cite{Scheffler2021IndirectionSS}. 
Thus, an in-order core can achieve at most \SI{11}{\percent} \gls{fpu} utilization, while a superscalar core would need at least nine issue slots to keep a single \gls{fpu} busy, even when ignoring dependencies and control bottlenecks. 
In both cases, the main efficiency bottleneck is the \emph{indirection} \verb|b[A_idcs[j]]|, which requires half of the non-compute instructions and is also needed for sparse-dense addition.

Now, consider the multiplication of \verb|A| with a sparse vector \verb|b| stored in the common \gls{csf} format. The \emph{sparse-sparse} dot product \verb|sV×sV| computed for each result element can be implemented as shown in \Cref{lst:intro_c}. %
We \emph{intersect} the nonzero indices in the two vectors since only positions where both vectors are nonzero contribute to the result.
Compared to \verb|sV×dV|, the control overhead is even higher and dynamically depends on the sparsity patterns of both operands. Sparse-sparse tensor addition needs yet another approach: we must form a \emph{union} of indices, since a nonzero in either operand produces a  nonzero in the result.

\begin{listing}[!t]
  \subfloat[(\alph{sublisting})~Multiplication of a sparse matrix \texttt{A} in \gls{csr} format with a dense vector \texttt{b} (\texttt{sM×dV}), iterating sparse-dense dot products (\texttt{sV×dV}).]{
    \label{lst:intro_a}
    \includegraphics[keepaspectratio]{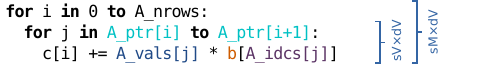}
  }\hfill
  \subfloat[(\alph{sublisting})~Dot product of two sparse vectors \texttt{a} and \texttt{b} in \gls{csf} format (\texttt{sV×sV}).]{
    \label{lst:intro_c}
    \includegraphics[keepaspectratio]{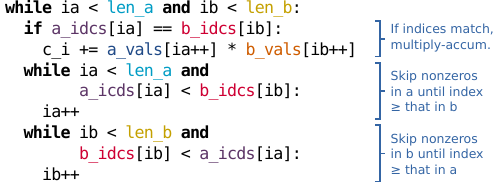}
  }
\caption{Example sparse-dense and sparse-sparse product kernels.}
\label{lst:intro_examples}
\end{listing}

Sparse compute is essential in many applications~\cite{Zhang2015ASO}. The SuiteSparse matrix collection~\cite{Davis2011TheUO} hosts sparse matrices from numerous real-world problems in domains including computational physics, mathematics, economics, and biology. 
In \gls{ml}, sparsification can significantly reduce the operations and memory required for a given inference accuracy \cite{Hoefler2021SparsityID}; while some approaches target sparse-dense compute by sparsifying only weights, others also exploit activation sparsity.
Graphs, commonly represented as highly sparse matrices, are operands in both sparse-dense and sparse-sparse workloads such as PageRank \cite{Page1999ThePC} and triangle counting \cite{Rao2022SparseCoreSI}.

Established superscalar out-of-order architectures struggle with the high control overhead of sparse compute:~\revone{a Fujitsu A64FX running \texttt{sM×dV} with a highly optimized format~\cite{Alappat2020PerformanceMO} achieves at most \SI{130.9}{\Gflops} or \SI{4.7}{\percent} of its peak compute.}
While recent CPU, GPU, and accelerator works improve sparse workload performance, they are limited by their lack of generality or large microarchitectural impact.
Most general-purpose \gls{isa} extensions consider only one-sided sparsity \cite{Talati2021ProdigyIT, Wang2019StreambasedMA}, and those targeting sparse-sparse workloads incur large area impacts and cannot efficiently handle the former \cite{Rao2022SparseCoreSI}. 
Sparse \gls{ml} solutions often rely on custom, domain-specific formats and low or structured sparsity \cite{Wang2021DualsideST, Gondimalla2019SparTenAS}, while more general sparse \gls{la} accelerators \cite{Hegde2019ExTensorAA, Dadu2019TowardsGP} incur significant silicon area impacts compared to their host systems and other solutions.

\glsunset{sssr}
To address these shortcomings, we present \emph{sparse stream semantic registers} (SSSRs), a modular \gls{isa} extension significantly accelerating and improving \gls{fpu} utilization on both sparse-dense \emph{and} sparse-sparse \gls{la} \emph{without} major microarchitectural impact.
\Glspl{sssr} %
extend \glspl{ssr}~\cite{Schuiki2021StreamSR}, which enable near-complete \gls{fpu} utilization in in-order cores for dense linear algebra and other data-oblivious workloads\footnote{A workload whose memory access patterns are input-invariant.}~\cite{Ramachandran2020DataOA} by mapping memory streams with regular address patterns directly to architectural registers.
We extend \glspl{ssr} to accelerate the \emph{indirection}, \emph{intersection}, and \emph{union} operations necessary for general sparse \gls{la}, incurring only a small increase in overall hardware complexity while significantly improving sparse performance. For example, \glspl{sssr} allow us to write both \verb|sV×dV| and \verb|sV×sV| in \riscv{} assembly as in \Cref{lst:intro_d}, where the registers \verb|ft0| and \verb|ft1| have stream semantics.
Since all index processing and comparison is handled in hardware, the loop body consists of only the useful \verb|fmadd.d| instruction, which we iterate over using a hardware loop.

We evaluate \glspl{sssr} by integrating them into the RISC-V Snitch \cite{Zaruba2021SnitchAT} core and compute cluster, in which we evaluate their benefits on various sparse \gls{la} workloads using generated vectors and real-world matrices.
On a single core, \gls{sssr} indirection, intersection, and union enable speedups of up to 7.0\x, 7.7\x, and 9.8\x~over an optimized RISC-V baseline on sparse-dense multiply, sparse-sparse multiply, and sparse-sparse addition, respectively, with peak indirection \gls{fpu} utilizations of up to \SI{80}{\percent}\footnote{Full \gls{fpu} utilization would require separate index memory ports, which we forego to minimize our architectural impact (see  \cref{sec:ext_ssrs}).}. On an eight-core cluster, sparse-dense \verb|sMxdV| using \glspl{sssr} is up to \revone{4.9}\x~faster and \revonemod{2.9}\x~more energy efficient, while sparse-sparse \verb|sMxsV| up to \revonemod{5.9}\x~faster and \revonemod{3.0}\x~more energy efficient. \glspl{sssr} incur only 1.8\% in additional area in an eight-core cluster (\SI{128}{\kibi\byte} data memory and \SI{8}{\kibi\byte} instruction cache) over regular \glspl{ssr}.
\begin{listing}[!t]
\includegraphics[keepaspectratio]{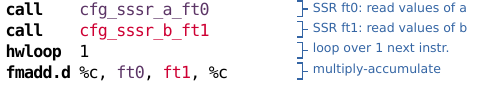}
\caption{\texttt{sV×dV} and \texttt{sV×sV} in RISC-V assembly using \glspl{sssr} and a hardware loop. The kernels differ only in the used \gls{sssr} configuration.}
\label{lst:intro_d}
\end{listing}
The synthesizable \gls{rtl} description for \glspl{sssr} and their integration into Snitch are available free and open-source\footnote{\url{https://github.com/pulp-platform/snitch}: \gls{sssr} IP in {hw/ip/}\\{snitch\_ssr} and integration in {hw/ip/snitch\_cluster}.}. 

\glspl{sssr} extend our prior work on \emph{indirection \glspl{ssr}} accelerating sparse-dense tensor products \cite{Scheffler2021IndirectionSS}. We include these contributions in this work to provide a holistic, coherent description of our \gls{ssr} sparsity extensions. All contributions regarding sparse-sparse \gls{la}, most parameterizability work, and many minor improvements are original to this work, and we updated all existing evaluation to a newer version of Snitch \cite{Zaruba2021SnitchAT} and a new technology node.
In detail, our contributions are:

\begin{figure*}[t]
  \setkeys{Gin}{height=7.6cm}
  \subfloat[Indirection address generator with\\datapath for each mode differentiated.]{
    \label{fig:arch_addr_gen}
    \includegraphics{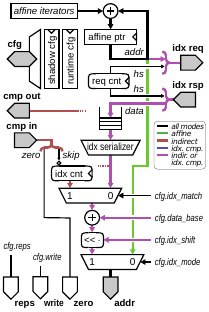}
  }\hfill
  \subfloat[Indirection and egress \glspl{ssr} with their new address generators and  preserved data mover in gray.]{
    \label{fig:arch_ssrs}
    \includegraphics{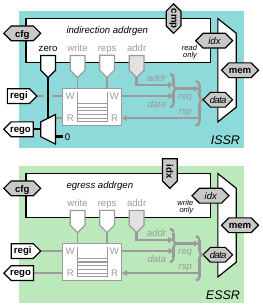}
  }\hfill
  \subfloat[Sparse \gls{ssr} streamer with two ISSRs, one ESSR, and index comparator between them.] {
    \label{fig:arch_sssr_streamer}
    \includegraphics{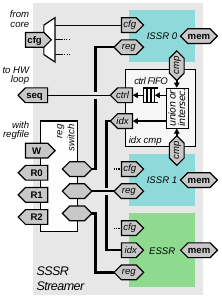}
    }
  \caption{Sparse SSR extension architecture, showcasing address generation, data movers, \gls{sssr} streamer, and inter-\gls{ssr} index comparison.}
  \label{fig:arch_overview}
\end{figure*}

\begin{enumerate}
    \item A lightweight, modular extension to \glspl{ssr} handling streaming indirection, intersection, and union in hardware to efficiently handle sparse general \gls{la} in single-issue cores, with a complete open-source hardware implementation in an open RISC-V core and multi-core cluster. (\Cref{sec:arch}).
    \item A programming model for \glspl{sssr}, efficient sparse-dense and sparse-sparse \gls{la} kernels operating on vectors and matrices, and further applications for our hardware (\Cref{sec:prog}).
    \item Significant performance and energy efficiency benefits on single cores ($\leq7.0,7.7,9.8$\x) and eight-core clusters \revone{with realistic interconnect and main memory models} ($\leq\revone{4.9},\revonemod{5.9}$\x~and $\revonemod{2.9},\revonemod{3.0}$\x)\revone{, incurring} a minimal \SI{1.8}{\percent} impact on cluster area (\Cref{sec:evaluation}).
    \item A comparison to state-of-the-art CPU, GPU, and accelerator approaches, achieving higher generality, a lower area impact, and \revonemod{1.7}\x~higher parallel \verb|sM×dV| \gls{fpu} utilization than GPUs (\Cref{sec:relwork}).
\end{enumerate}

The paper is organized as follows: \Cref{sec:arch} describes our \gls{sssr} architecture and its integration into the Snitch core and cluster. \Cref{sec:prog} presents our \gls{sssr} programming model and sparse \gls{la} kernels. \Cref{sec:evaluation} evaluates the performance, timing, area, and energy efficiency benefits of \glspl{sssr}. Finally, \Cref{sec:relwork} discusses related CPU, GPU, and accelerator work and compares it to our approach.

\section{Architecture}\label{sec:arch}

Most of our hardware extensions are confined to the address generator inside each \gls{ssr}. We will first propose two new generator designs, the \emph{indirection} and \emph{egress} address generators, as well as two extended, backward-compatible \gls{ssr} hardware variants using them. We will then present the sparse \gls{ssr} streamer, which combines and connects the index datapaths of three extended \glspl{ssr} to enable index intersection and union. Finally, we will discuss the streamer's integration into the Snitch core and cluster~\cite{Zaruba2021SnitchAT} in which we evaluate our extensions. We note that \glspl{sssr} are not strictly dependent on a specific core or instruction set and can be used with any core, \gls{isa}, or hardware loop.

\subsection{Address Generators}
\label{sec:agens}

We designed two extended \gls{ssr} address generators to handle indexed streams. The \emph{indirection} generator can \emph{read} an index stream from memory to either generate indirect addresses or match read indices against those of another \gls{issr} for intersection or union. The \emph{egress} generator can \emph{write} an externally provided index stream to memory while emitting corresponding data addresses. \Cref{sec:sssr_streamer} explains how these features can be combined to enable end-to-end sparse-sparse index processing. 

\subsubsection{Indirection Address Generator}

\cref{fig:arch_addr_gen} shows the indirection address generator, highlighting the datapaths used by different generation modes. Like the original generator design, it exposes shadowed configuration registers to the host core, enabling the setup of a new stream while another is still running, and a control interface to the surrounding \gls{ssr} logic. The index stream is fetched through an additional read-only memory port.

All generation modes reuse the existing affine address generator with up to four nested levels, but for different purposes. During \emph{affine iteration}, it directly provides the address generator's output as shown. During \emph{indirection} or \emph{index matching}, %
it generates addresses for the words in memory containing the indices to be read; unlike fetching individual indices, this fully utilizes the memory bus. These words are fetched into a decoupling \gls{fifo} queue. To prevent downstream blocking or a queue overflow, an outstanding request counter limits inflight requests.

\glspl{sssr} can handle \revone{indices of any unsigned $2^n$\,\si{\byte} integer type} that fits onto the used memory bus (i.e., $8$, $16$, $32$, and \SI{64}{\bit} on Snitch). An \emph{index serializer} extracts indices of the configured size from the buffered index words. To increase flexibility and programmability, we support arbitrary index \revone{array} base alignment \revone{in memory}. In indirection mode, the serialized indices are then shifted by a small programmable offset, enabling power-of-two striding; this enables indirection into the upper axes of power-of-two-strided tensors without incurring a costly hardware multiplier. Finally, the shifted indices are added to the configured base address, resulting in the desired indirect address stream. 

During index matching, the indices are instead passed to an external matching unit, detailed in \cref{sec:sssr_streamer}, which compares the read indices with those in another \gls{ssr}'s generator streaming the other operand. Unlike in indirection, the emitted addresses simply count up from the configured data base address with a unit stride, since there is exactly one nonzero value per index.
However, depending on which index stream is ahead, the matching unit may instruct either generator to \emph{skip} addresses or \emph{insert} null data elements into the \gls{ssr} data stream, forming the \emph{intersection} or \emph{union} of the index streams.

\subsubsection{Egress Address Generator}

The \emph{egress} address generator is much simpler. It addresses the need to write back the indices of a joint stream alongside its data. For this purpose, it reads joint indices generated by an index matching unit through an additional port, passes them to an index \emph{coalescer}, and writes them back through a \emph{write-only} memory port, effectively reversing the indirection generator's index fetch datapath. It writes exactly one index per emitted address, but allows index writing to lead by a parameterized number of elements, decoupling the index and data streams and preventing unnecessary blocking.

Both extended generators are fully parameterized. The address and data widths, affine loop count, and index queue depth can all be adapted. Reduced internal index and address widths may be used for \revone{area-efficient designs operating} on smaller memories.

\begin{figure}[t]
    \centering
    \includegraphics[width=\linewidth]{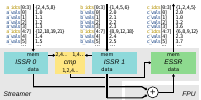}
    \caption{Example \gls{sssr} memory request sequences and index union dataflow for end-to-end \texttt{sV+sV} with 64-bit FP data and 16-bit indices.}
    \label{fig:arch_svPsv_example}
\end{figure}

\subsection{Sparse SSRs}
\label{sec:ext_ssrs}

\cref{fig:arch_ssrs} shows our extended \glspl{ssr} built around the proposed address generators, the \gls{issr} and \gls{essr}. Like the original \gls{ssr}, both provide a data queue decoupling the register and memory ports which is used in both read and write directions. 
Both expose their address generator's index matching ports, and the \gls{issr} adds some multiplexing logic to enable the indirection generator to inject zero data elements into read streams for stream union. %

In both the \gls{issr} and \gls{essr}, we chose to combine the index and data memory ports using a round-robin arbiter instead of requiring two memory ports. %
During affine streaming, this does not affect data throughput as the index port is unused. For both indirect and egress streaming, this tradeoff is motivated as follows: only one index word must be transmitted each $n$ data words, where $n$ is the number of indices per memory word. Thus, the peak achievable data mover utilization \revone{is} $\nicefrac{n}{n+1}$; on our evaluation platform with $64$-bit data, this corresponds to $67\%$, $80\%$, and $88\%$ for $32$-, $16$-, and $8$-bit indices, respectively. %
Alternatively, we could arbitrate memory accesses in the streamer or expose two core memory ports per \gls{ssr}, trading higher peak utilization on sparse streams for a significantly larger cluster interconnect. In this work, we consider area-efficient \gls{sssr} variants with one port, enabling the drop-in replacement of and direct comparisons to the existing single-port \glspl{ssr}.

\subsection{Sparse SSR Streamer}
\label{sec:sssr_streamer}

\cref{fig:arch_sssr_streamer} shows the \gls{sssr} \emph{streamer} in its default configuration. Like the original \gls{ssr} streamer, it combines all available \glspl{ssr} and provides shared configuration and register file interfaces to the host processor. A \emph{register switch} maps each \gls{ssr}'s data channel to a predefined register when enabled.

The \gls{sssr} streamer is fully parameterizable and may be configured to map any number of \glspl{ssr}, \glspl{issr}, or \glspl{essr} to any register indices. However, its optional inter-\gls{ssr} index stream join capabilities, if present, are subject to some configuration constraints. A single streamer provides at most \emph{one} index comparator; thus, only \emph{two} \glspl{issr} in a streamer may support index comparison between each other, and one optional third \gls{essr} may consume the joined index stream for writeback. Thus, the default configuration comprises two index-comparing \glspl{issr} and one \gls{essr}.

The index comparator %
determines which, if either, of the \gls{issr} index streams is ahead, and advances both as necessary to emit the union or intersection of indices. %
\Cref{fig:arch_svPsv_example} demonstrates on an \verb|sV+sV| example how the index comparator generates a union of \gls{issr} indices and forwards them to an \gls{essr} for writeback, while the host core's FPU performs the required additions.
The desired comparison mode is configured in and forwarded from the \glspl{issr}. %
Since the number of index-data pairs in a joint stream is unknown ahead of time, the comparator also exposes a generic \emph{stream control} queue, which provides a single bit indicating either that another element pair is available or that the joint data stream is complete. This queue can be read by a host core's branch \revone{unit} or hardware loop unit in a decoupled fashion to determine the end of a joint stream.

\begin{figure}[t]
    \centering
    \includegraphics[width=\linewidth]{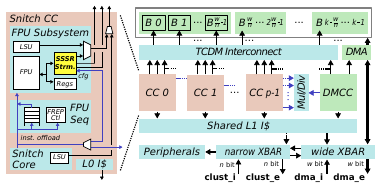}
    \caption{Snitch Cluster used in evaluation with SSSR streamer}
    \label{fig:arch_cluster}
\end{figure}

\begin{table}
\setlength{\tabcolsep}{0.6em}
\centering
\begin{tabular}{lllr}
\toprule
 & Parameter                         & Possible values                               & Used \\
 \midrule
$p$ & Worker core count              & $\mathbb{N}^{> 0}$                            & 8    \\
$n$ & Narrow (core, bank) width      & 32, 64                                        & 64   \\
$w$ & Wide (L1 I\$, DMA) width       & $\{n \leq 2^j \leq 1024~|~j \in \mathbb{N}\}$ & 512  \\
$k$ & Memory bank count              & multiples of~$\nicefrac{w}{n}$                & 32   \\
$D$ & TCDM size in \si{\kibi\byte}   & $\{2^j~|~j \in \mathbb{N}\}$                  & 128  \\
$I$ & L1 I\$ size in \si{\kibi\byte} & $\{2^j~|~j \in \mathbb{N}\}$                  & 8    \\
\bottomrule
\end{tabular}
\caption{Snitch cluster parameters and values used in evaluation.}
\label{tbl:arch_cluster}
\end{table}

\subsection{Core and Cluster Integration}
\label{sec:arch_cc_clust}

\cref{fig:arch_cluster} shows the integration of the default \gls{sssr} streamer into the Snitch \gls{cc} and cluster~\cite{Zaruba2021SnitchAT} used in evaluation. Like the previously integrated \gls{ssr} streamer, it is multiplexed with the double-precision \gls{fpu}'s register file.
The \glspl{cc} preserve their existing memory topology, combining  the core, FPU, and first \gls{issr} into one memory port and providing exclusive ports to the second \gls{issr} and \gls{essr}.
We also extend the existing \gls{frep} hardware loop with a new mode using our stream control interface to issue one iteration per joint stream element.

\Cref{tbl:arch_cluster} lists the essential design parameters and their possible values for the fully configurable cluster. All cores share an L1 instruction cache and an integer multiply-divide unit. A \revone{\gls{dma} engine connected to a wide memory bus} enables high-bandwidth data transfers between the \gls{tcdm} and main memory. It is controlled by a lightweight \gls{dmcc} which can also be used for cluster management and coordination.

\section{Programming and Applications}\label{sec:prog}

The \glspl{sssr} are configured by the host core using a register interface accessed through custom instructions. This enables fast job setups: in our kernels, it takes at most 10 cycles to set up and launch new jobs for all three \glspl{sssr} from a cold state. In our configuration, \glspl{issr} 0 and 1 map to temporary \gls{fpu} registers \verb|ft0| and \verb|ft1| and the \gls{essr} to \verb|ft2|. The redirection of register accesses to \glspl{ssr} is toggled through a common control register~\cite{Schuiki2021StreamSR}. We focus on accelerating sparse \gls{la} at the instruction level
by creating a library of hand-optimized operators, as is commonly done in high-performance \gls{la} packages. Future work will focus on extending compilers to automatically infer indirections, intersections, and unions in high-level code and map them to \glspl{sssr}.

\glsunset{csc}
\subsection{Accelerable Formats and Operations}
\label{sec:acc_fmt_prob}

While we explore further workloads in \Cref{sec:furth_app}, our main motivation is accelerating general sparse \gls{la} on a set of established tensor formats. 
Unlike approaches focused on low sparsities \cite{nvidia2020a100, Wang2021DualsideST, Gondimalla2019SparTenAS}, \glspl{sssr} target flexibility and scalability to achieve notable speedups across a wider sparsity range (see \cref{sec:eval_cc}).
\glspl{sssr} support iterating along the major axis of any tensor format where said axis is given by two arrays: a \emph{value} array storing nonzero values and an \emph{index} array storing their positions. We call this pair of arrays a sparse \emph{fiber}.
Fiber-based tensor formats include the common \gls{csr}~\cite{eisenstat1977csr} and \gls{csc}~\cite{duff1989csc} matrix formats, the generalized \gls{csf}~\cite{Smith2015TensormatrixPW} tensor format, and many format variations that slice data, pack it into blocks, or further compress upper tensor dimensions.
The index shifter discussed in \Cref{sec:agens} also enables iterating on higher-level axes if all axes beneath it have power-of-two dimensions, which can be ensured through data chunking in practice. %
Indices can be of any unsigned $2^n$\,\si{\byte} \revone{integer} type that fits onto the memory bus.%

\revone{The data width of \glspl{sssr} is fixed to that of the host processor, but \gls{simd} computations on blocked formats like BCSR \cite{Pinar1999ImprovingPO} are trivially supported through zero padding.
For example, the 64-bit \glspl{fpu} in Snitch can leverage their 2- and 4-way \gls{simd} capabilities on 32- and 16-bit formats when computing on blocks of two or four elements, respectively. 
We can operate on nonzero \gls{simd} blocks the same way as on single double-precision elements; this substitution is orthogonal to the use of \glspl{sssr} and simply consolidates narrow nonzero elements into 64-bit nonzero \gls{simd} blocks. Therefore, \glspl{sssr} should enable similar peak speedups and somewhat reduced speedups for a given nonzero pattern, as proportionally more nonzero elements must pack together to overcome \gls{sssr} setup costs.}

\glspl{sssr} are not bound to specific sparsity regimes. They enable significant efficiency gains on sparse-sparse, sparse-dense, and even dense problems through \gls{ssr} backward-compatibility. The supported formats have no structural sparsity requirements and scale well to extreme sparsities. Furthermore, \glspl{sssr} do not impose a specific higher-order dataflow: for example, sparse-sparse matrix multiplication (\verb|sM×sM|) may be done in an inner-product, outer-product, or row-wise fashion to best suit the application and data.

To optimize area and leakage, the \gls{sssr} streamer parameterization can be tailored to specific application requirements. For efficient sparse-dense multiplication, one \gls{issr} and one \gls{ssr} are sufficient, with an additional \gls{issr} needed for addition. For efficient sparse-sparse multiplication, two \glspl{issr} suffice, with an additional \gls{essr} incurred for addition.

\subsection{Sparse Linear Algebra Kernels}
\label{sec:acc_kernels}

To demonstrate the versatility of our extensions and evaluate their benefits, we create a set of sparse-dense and sparse-sparse linear algebra kernels for the \riscv{} Snitch system extended with \glspl{sssr}. The kernels operate on \gls{csr} matrices and \gls{csf} vectors. Each is implemented for 8-, 16-, and 32-\si{\bit} index types and, except for intersection-based kernels, in three variants:

\begin{itemize}
    \item \textsc{base}: Stock \riscv{} optimized baseline
    \item \textsc{ssr}: \riscv{} extended with \gls{frep} loop and \glspl{ssr}
    \item \textsc{sssr}: \riscv{} extended with \gls{frep} loop and \glspl{sssr}
\end{itemize}

To obtain the \textsc{base} kernels, we compile high-level C code with \verb|clang -Ofast| and hand-optimize the resulting assembly for Snitch \glspl{cc} through instruction reordering and loop unrolling to minimize \revone{memory or dependency} stalls.
The \textsc{ssr} kernels further improve performance by mapping regular, affine value streams such as sparse value arrays to \glspl{ssr}, but without using our sparsity extensions. We note that this excludes kernels using intersection, since regular \glspl{ssr} cannot further accelerate conditional stream loads.
Finally, we implemented all kernels as \textsc{sssr} variants fully leveraging our extensions, which we will discuss in detail.

\subsubsection{Sparse-dense Kernels}

\begin{listing}[!t]
\includegraphics{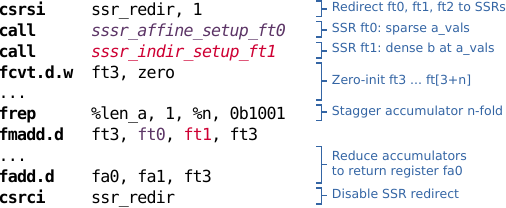}
\caption{\textsc{sssr} variant of \texttt{sV×dV} kernel: \texttt{ft0} streams the values of sparse vector \emph{a} and \texttt{ft1} streams indirected values from dense vector \emph{b}.}
\label{lst:prog_svXdv_kernel}
\end{listing}

\topar{sV×dV} \cref{lst:prog_svXdv_kernel} shows our \textsc{sssr} \verb|sV×dV| kernel. Versions for different index sizes differ only in the indirection configuration for \texttt{ft1}. Our goal is to continuously issue the \gls{mac} instruction \verb|fmadd.d| with minimal overhead to maximize \gls{fpu} utilization. We achieve this in three phases:

\begin{enumerate}
    \item \emph{Setup}: We first enable register redirection to \glspl{ssr}. We then set up \gls{issr} \verb|ft0| to linearly stream the sparse vector \verb|a|'s values in affine mode and \gls{issr} \verb|ft1| to stream the dense vector \verb|b| indirected at \verb|a|'s indices. Finally, we zero-initialize a contiguous block of accumulator registers starting at \verb|ft3|.
    \item \emph{Compute}: We use \gls{frep} to iterate the useful \verb|fmadd.d| instruction, streaming both multiplicands from the previously configured \glspl{issr}. To prevent stalls caused by register dependencies on the accumulated sum, we use \gls{frep}'s existing register staggering feature to increment the accumulator register index on each iteration, maintaining several partial sums in our contiguous block of registers; this is described in detail by Zaruba et al.~\cite{Zaruba2021SnitchAT}. Due to the limited peak throughput our \glspl{issr} with one memory port impose, the larger the index type, the fewer accumulators are needed to hide these  stalls. 
    \item \emph{Teardown}: We reduce our accumulators into the return register \verb|fa0| and disable \gls{ssr} redirection.
\end{enumerate}

Since the compute and teardown phases contain only \gls{fpu} instructions, the core is free to continue execution while the \gls{fpu} is kept busy by \gls{frep}; this also enables the seamless setup of new shadowed \gls{ssr} jobs. Synchronization between the core and \gls{fpu} can be enforced whenever necessary.

\topar{sM×dV} Instead of simply iterating our \verb|sV×dV| kernel for each matrix row, we further optimize \verb|sM×dV| to reduce setup and reduction overheads and maximize FPU utilization. We stream the entire matrix fiber in single \gls{ssr} and \gls{issr} jobs, significantly reducing our setup overhead. Furthermore, we unroll the first few \verb|fmadd| in each row, adding branches to shorter reductions, and issue an \gls{frep} loop and a full reduction only when necessary to accelerate short rows.

We use 32-bit row pointers in all variants to maximize row scaling. While we default on multiplying a \gls{csr} matrix from the left, our kernels provide runtime parameters to use different power-of-two vector and arbitrary result strides; this enables multiplying a \gls{csr} or \gls{csc} matrix with any power-of-two-strided dense tensor axis from either side.

\topar{sM×dM} We multiply a \gls{csr} matrix with a power-of-two-column row-major matrix. We simply iterate our \verb|sMxdV| kernel here, as the additional overhead of iterating over dense data in a third-order loop is negligible (see \cref{sec:eval_cc_sd}). We again provide parameters for custom dense matrix and result strides, enabling multiplication of  \gls{csr} or \gls{csc} matrices with row- and column-major matrices from either side. The only restriction is that our hardware index shifter requires the dense axis we indirect into to be power-of-two-strided; in practice, this can be accommodated by tiling matrices using the cluster \gls{dma}'s strided transfers.

\topar{sV+dV} We assume that the result is accumulated onto the dense vector; if necessary, a copy can be created using \glspl{ssr} beforehand. We configure \verb|ft2| to stream the sparse vector values, \verb|ft0| to gather the dense addends, and \verb|ft2| to scatter sum elements back into the dense vector, and use \gls{frep} to execute \verb|fadd.d ft1, ft0, ft2| once for each sparse value. Either vector could also be scaled by a constant with the same performance by iterating an \verb|fmadd| instead. This element-wise addition can be extended to fiber-based tensors of \emph{any} shape through higher-level iteration.

\topar{sV$\odot$dV} Since one operand is dense, the result indices are the same as the sparse operand indices; they can optionally be copied with \glspl{ssr} if needed. We multiply each sparse value with a gathered dense co-operand as for \verb|sV+dV|, but instead linearly write out the results to yield the corresponding result value array. Again, this element-wise multiplication can be iterated to accommodate tensors of any shape.

\subsubsection{Sparse-sparse Kernels}

\begin{listing}[!t]
\includegraphics{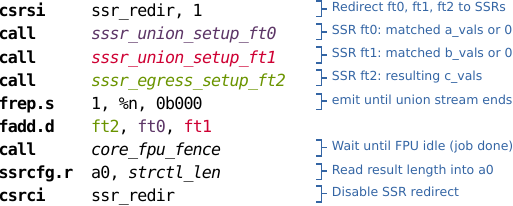}
\caption{\textsc{sssr} variant of \texttt{sV+sV} kernel: \texttt{ft0} and \texttt{ft1} stream in sparse operands \emph{a} and \emph{b} and \texttt{ft2} streams out sparse result \emph{c}.}
\label{lst:prog_svPsv_kernel}
\end{listing}

\topar{sV×sV} Since both indirection and intersection are fully handled inside \glspl{sssr}, the \textsc{sssr} \verb|sV×sV| kernel is identical to its \verb|sV×dV| counterpart except for the \gls{sssr} and \gls{frep} configurations used. We configure \verb|ft0| and \verb|ft1| to stream the fibers of each operand and activate intersection such that only value pairs with matching indices are streamed to the \gls{fpu}. Since the number of emitted pairs is unknown ahead of time in intersection, we use our stream-controlled \gls{frep} variant (\verb|frep.s|) to issue one \verb|fmadd.d| for each. 

\topar{sM×sV} We iterate our \verb|sV×sV| code for each result element. Since our intersection approach requires increasing indices in fibers, we launch new \gls{sssr} jobs for each row unlike for \verb|sM×sV|. However, we can hide some of this configuration overhead through the decoupling of core and \gls{fpu} subsystem and the shadowed \gls{sssr} configuration interface.

\topar{sM×sM} Three sparse-sparse matrix multiply dataflows, all accelerable with \glspl{sssr}, are commonly differentiated:

\begin{itemize}
    \item \emph{inner}: compute output element-wise by computing inner products (\verb|sV×sV|) for each row-column pair. 
    \item \emph{row-wise}: compute output row-wise by accumulating  sparse rows of left matrix (\verb|sV+sV|) scaled with nonzeros in each row of right matrix. 
    \item \emph{outer}: compute outer product for each row-column pair (dense vector scaling), then accumulate all partial sparse result matrices (\verb|sM+sM)|.
\end{itemize}

Each method has a different iteration order, memory footprint, communication overhead, and data reuse, making each attractive for different applications and sparsity regimes. \glspl{sssr} can accelerate all three methods by accelerating the subkernels pointed out above. Here, we limit ourselves to implementing a \gls{csr}×\gls{csc} \emph{inner} kernel iterating our \verb|sM×sV| kernel.

\topar{sV+sV} As shown in \cref{lst:prog_svPsv_kernel}, we read two sparse vectors from \glspl{issr} \verb|ft0| and \verb|ft1| and write their sparse sum to \gls{essr} \verb|ft2|. The index comparator performs index \emph{union} so that \verb|ft0| and \verb|ft1| emit either an element value or \emph{zero} if an index only occurs in the co-operand. This allows us to simply iterate \verb|fadd.d| as shown, with our stream-controlled \verb|frep.s| ensuring the correct iteration count. Since \verb|ft2| is configured as an egress, one compared index will be written to memory for each computed sum value, resulting in the sum index vector. We read the result length from an \gls{essr} configuration register after the job is done, which we ensure by synchronizing the core and \gls{fpu} pipelines. When iterating this kernel to add higher-dimensional sparse tensors (e.g., \verb|sM+sM|), we can defer this read to overlap computation with configuration as for \verb|sM×sV|.

\topar{sV$\odot$sV} This kernel is almost identical to \verb|sV+sV|; we instead configure the index comparator for \emph{intersection} and iterate \verb|fmul.d|. The same insights on higher-dimensional tensors and control-compute overlap also apply here.

\subsection{Further SSSR Applications}
\label{sec:furth_app}

\glspl{sssr} target general sparse \gls{la}, accelerating the many applications built on it including \gls{fem} simulations, sparse and graph neural networks, and graph workloads like PageRank \cite{Page1999ThePC}. However, the indirection, intersection, and union they accelerate are \emph{general-purpose} operations that underlie many more irregular workloads, some of which we want to \revone{discuss} here:

\topar{Codebook decoding} \glspl{issr} can stream compressed data stored as index arrays pointing into smaller arrays of repeated values. Such \emph{codebooks} could be used to efficiently store quantized deep learning parameters~\cite{Dave2021HardwareAO}, image channels~\cite{marcellin2002jpeg}, and nonzero values of sparse tensors. Each \gls{issr} in a streamer can independently stream codebooked vectors. Two \glspl{issr} can handle sparse-dense \gls{la} with codebooked nonzeros (one gathering dense operands, the other nonzeros) with similar performance as our sparse-dense kernels, significantly reducing the size of compressible tensors.

\topar{Stencil codes} Iterative stencil codes process data on $n$-dimensional grids by accessing arrays in fixed, but irregular patterns relative to each grid point. Examples include partial differential equations and image processing \cite{Roth1997CompilingSI}. \glspl{issr} can accelerate stencils by storing them as index arrays and streaming them for each grid point, using its offset as a base address. The same approach can also accelerate sparse convolutions without im2col-like preprocessing.

\topar{Graph pattern matching} Intersection between the sparse adjacency vectors of graph nodes can be leveraged to identify and count subgraph embeddings and assess the similarity of graphs \cite{Rao2022SparseCoreSI}. Thus, \glspl{sssr} may be used to accelerate graph pattern matching workloads in fields like computer vision and drug discovery.

\topar{Scatter-gather data transforms} \glspl{issr} enable a streaming variant of \emph{scatter-gather}. Thus, they can accelerate any existing scatter-gather-based transforms rearranging data, including sparse matrix transpose~\cite{Stathis2004SparseMT}, parallel radix sort\cite{Zagha1991RadixSF}, or densifying sparse tensors by scattering their nonzeros.

\section{Evaluation}
\label{sec:evaluation}

We will first evaluate the performance benefits of \glspl{sssr} on our \gls{la} kernels on a single Snitch \gls{cc} and in parallel cluster \verb|sM×dV| and \verb|sM×sV| implementations \revone{backed by a realistic \gls{dram} simulator}. Next, we will discuss the area and timing impacts of different hardware configurations of our streamer. Finally, we will investigate the energy efficiency benefits of our extensions.

\revone{Like the baseline Snitch platform we integrate them into, our hardware extensions are designed and modeled as parametric, fully synthesizable SystemVerilog \gls{rtl} hardware descriptions. This allows us to simulate both the baseline and our extended systems with full cycle accuracy and visibility, enabling the precise tracing and utilization tracking of all components. While we could use any SystemVerilog-capable simulator, we use \emph{Questa Advanced Simulator} here. Our implementations used to evaluate area, timing, and power use the same \gls{rtl} descriptions we simulate as inputs. Hardware resources external to the simulated \gls{cc} or cluster are modeled behaviorally.} Unless noted otherwise, the evaluated cluster and \gls{cc} use default \gls{sssr} streamers and are parameterized as specified in \cref{tbl:arch_cluster}.

For all experiments, dense test tensors are obtained by sampling normally distributed values and sparse vectors are generated for a given nonzero count and dimension with normally distributed values and uniformly distributed indices. The sparse matrices used are all real-world-problem matrices from the SuiteSparse Matrix Collection \cite{Davis2011TheUO}; they  cover various problem domains and aspect ratios and have~2\si{\kilo\relax} to 3.2\si{\kilo\relax} columns and 2.8\si{\kilo\relax} to 543\si{\kilo\relax} nonzeros.

\subsection{Single Core Performance}
\label{sec:eval_cc}

\begin{figure*}[t]
  \setkeys{Gin}{height=4.51cm}
  \subfloat[\gls{cc} \texttt{sV×dV} \gls{fpu} utilizations vs. sparse vector nonzeros (dashed: without reductions).]{
    \label{fig:res_cc_svXdv_util}
    \includegraphics{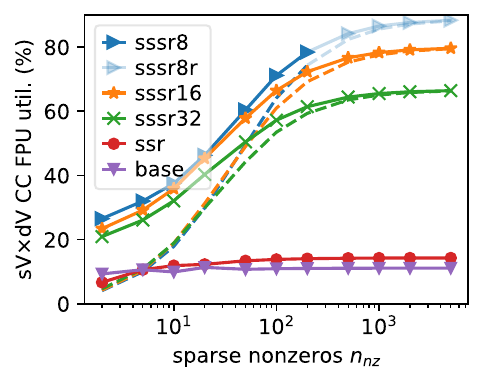}
  }\hfill
  \subfloat[\gls{cc} \texttt{sV+dV} \gls{fpu} utilizations vs. sparse vector nonzeros (no reductions needed).]{
    \label{fig:res_cc_svPdv_util}
    \includegraphics{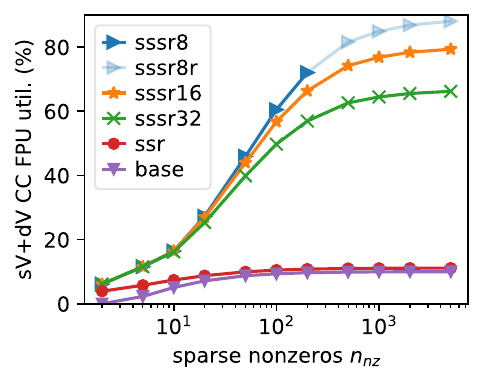}
  }\hfill
   \subfloat[\gls{cc} \texttt{sM×dV} speedups of \textsc{ssr} and \textsc{sssr} over \textsc{base} kernels for selected matrices.] {
    \label{fig:res_cc_smXdv_speedup}
    \hspace{0.5em}
    \includegraphics[height=4.51cm]{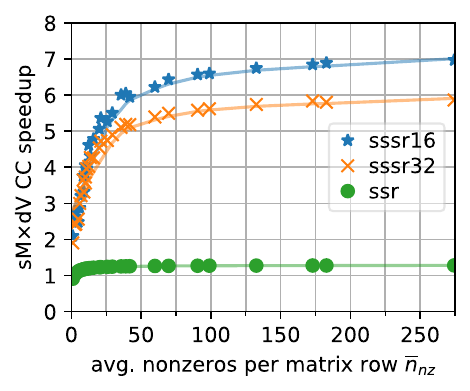}
  }\\
  \subfloat[\gls{cc} \texttt{sV×sV} speedups of \textsc{sssr} over \textsc{base} kernel. Both vectors have dense size 60\si{\kilo\relax}.] {
    \label{fig:res_cc_svXsv_speedup}
    \includegraphics{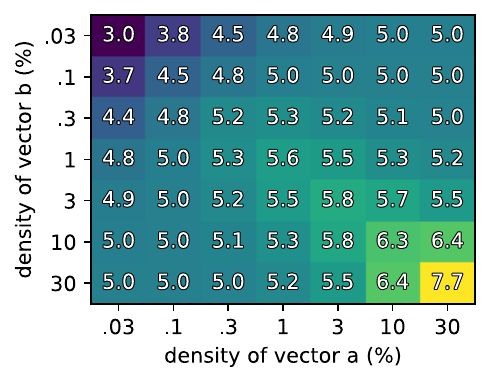}
  }\hfill
  \subfloat[\gls{cc} \texttt{sV+sV} speedups of \textsc{sssr} over \textsc{base} kernel. Both vectors have dense size 60\si{\kilo\relax}.] {
    \label{fig:res_cc_svPsv_speedup}
    \includegraphics{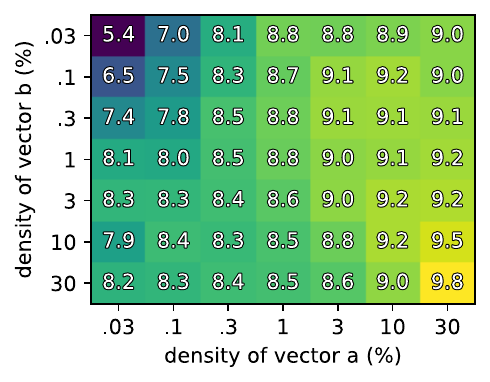}
  }\hfill
  \subfloat[\gls{cc} \texttt{sM×sV} speedups of \textsc{sssr} over \textsc{base} kernel for selected matrices.] {
    \label{fig:res_cc_smXsv_speedup}
    \includegraphics{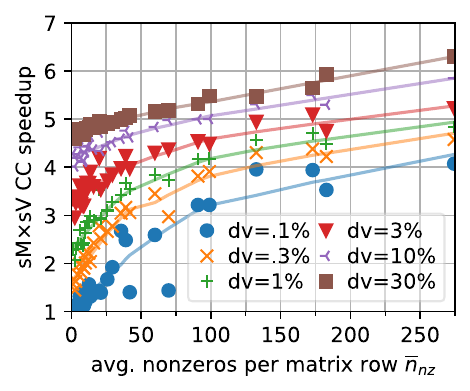}
  }\\
  \caption{Single-\gls{cc} performance results for selected kernels leveraging indirection, intersection, and union. \ref{fig:res_cc_svXdv_util} and \ref{fig:res_cc_svPdv_util} include results assuming 8-bit indices occur multiple times (\texttt{sssr8r}).  The overlaid trend lines in \ref{fig:res_cc_smXdv_speedup} and \ref{fig:res_cc_smXsv_speedup} are fitted using \gls{loess}.}
  \label{fig:res_cc}
\end{figure*}

To evaluate the performance benefits of \glspl{sssr} without the influence of other system components, we first evaluate our \gls{la} kernels in \gls{rtl} simulation of a single \gls{cc} by connecting it to an exclusive instruction cache and \revone{exclusive} three-port data memory. These behave similarly to the shared instruction cache and \gls{tcdm} channels in a cluster, respectively, except for additional cache misses and bank conflicts whose effects we include and discuss in \cref{sec:results_cl}.

\subsubsection{Sparse-dense Kernels}
\label{sec:eval_cc_sd}

\topar{sV×dV} \cref{fig:res_cc_svXdv_util} shows the \gls{fpu} utilization for our \verb|sV×dV| kernels for varying sparse vector nonzero counts $n_{nz}$. We note that the kernel runtimes do not depend on the dense vector's length as long as it fits into the \gls{tcdm}. Non-\textsc{sssr} variants perform identically for all index sizes as a RISC-V load of any size incurs one instruction. \textsc{sssr} kernel \gls{fpu} utilizations are shown with and without accumulator reductions (dashed). Since an 8-bit-indexed \gls{csf} vector can hold at most 256 nonzeros, we also show results for 8-bit-index indirection with repeated indices (\texttt{sssr8r}).

All three \textsc{sssr} index variants significantly outperform the \textsc{base} and \textsc{ssr} kernels, with 16- and 32-\si{\bit} variants approaching their arbitration-imposed utilization limits of  \SI{67}{\percent} and \SI{80}{\percent} and achieving 4.7\x~and 5.6\x~higher \gls{fpu} utilization than \textsc{ssr}. The 8-bit-indexed kernel reaches up to 5.8\x~higher utilization, and 8-bit indirection with repeated indices approaches its arbitration limit of \SI{88}{\percent}. We note that \textsc{base} and \textsc{ssr} also approach their issue-bound utilization limits of one \gls{mac} every nine and seven cycles, respectively.

Because they process nonzeros much faster, the \textsc{sssr} kernels need a significantly higher $n_{nz}$ to fully overcome their setup and reduction overheads. For few nonzeros, their utilization without reductions is even lower than for \textsc{base} and \textsc{ssr}, motivating our row unrolling in \textsc{sssr} \verb|sM×dV| kernels. Since \textsc{sssr} kernels with smaller indices need more accumulators to sustain their peak utilization, they outperform their larger-index counterparts only for higher $n_{nz}$.

\topar{sV+dV}  \cref{fig:res_cc_svPdv_util} shows the FPU utilization for our \verb|sV+dV| kernels. Here, the result is \emph{accumulated} onto the dense vector. We observe similar trends as for \verb|sV×dV| with two key differences. Since a result value is written back for each nonzero, \textsc{base} and \textsc{ssr} peak utilizations further decrease to one \gls{mac} every ten and nine cycles. For the same reason, no more reduction is necessary in \textsc{sssr} kernels as results are immediately written back by streaming indirection using the second \gls{issr}. The \texttt{sssr} kernels using 32-bit indices, 16-bit indices, and 8-bit indices with reuse still approach their \revone{arbitration-imposed} limits of 67, 80, and \SI{88}{\percent}. 

\topar{sM×dV} \cref{fig:res_cc_smXdv_speedup} shows the speedups of \textsc{sssr} \texttt{sM×dV} kernels over \textsc{base} against the average nonzeros per matrix row $\overline{n}_{nz}$, reflecting the \glspl{mac} performed per row. We assume here that the \gls{tcdm} is large enough to store the full matrix. We do not evaluate 8-bit indices as they are not large enough to index the chosen matrices' columns.

As for \texttt{sV×dV}, our 16-bit and 32-bit \textsc{sssr} kernel speedups over \textsc{base} approach their theoretical limits and reach up to 5.9\x~and 7.0\x, requiring a higher $\overline{n}_{nz}$ than \textsc{base} and \textsc{ssr} to compensate their row iteration and reduction overheads. \gls{fpu} utilizations reach up to \SI{66}{\percent} and \SI{79}{\percent}, respectively. Again, the 16-bit \textsc{sssr} kernel outperforms the 32-bit variant only once $\overline{n}_{nz} \apprge 20$ due to its longer reduction sequence.

\topar{sM×dM} Even in edge cases, speedups and \gls{fpu} utilizations are nearly identical to the \verb|sM×dV| kernels we iterate: for the tiny sparse matrix \emph{Ragusa18} with only 64 nonzeros, \gls{fpu} utilization changes by only \SI{0.12}{\percent} when multiplying it with a two-column dense matrix instead of a dense vector.

\begin{revonefig*}[ht!]
  \vspace{-0.6em}%
  \hspace{0.75ex}%
  \setkeys{Gin}{height=4.41cm}%
  \subfloat[Cluster \texttt{sM×dV} speedups of \textsc{sssr} over \textsc{base} for 16-bit indices and selected matrices.]{%
    \centering%
    \includegraphics{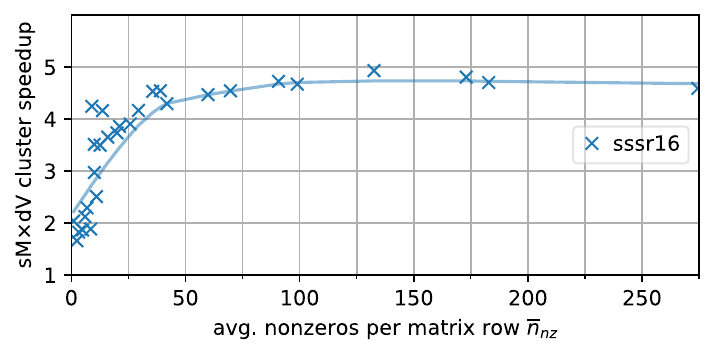}
    \label{fig:res_cl_csrmv_speedup}
  }\hfill%
  \subfloat[Cluster \texttt{sM×sV} speedups of \textsc{sssr} over \textsc{base} for 16-bit indices, varying vector densities, and selected matrices.]{%
    \centering%
    \includegraphics{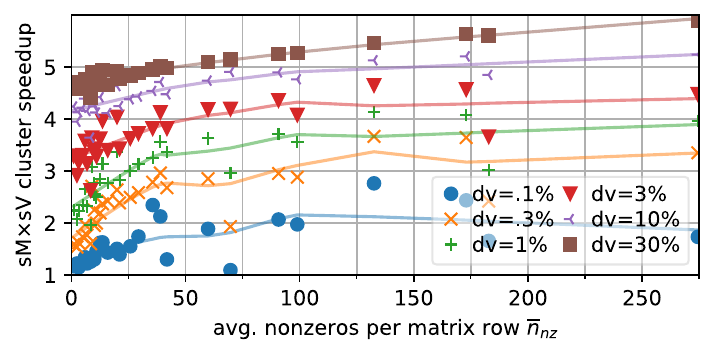}
    \label{fig:res_cl_csrmspv_speedup}
  }%
  \hspace{75ex}%
  \\
  \caption{Eight-core cluster performance results for selected \gls{la} kernel scaleouts. Trend lines are fitted using \gls{loess}.}
  \label{fig:res_cl}
\end{revonefig*}

\subsubsection{Sparse-sparse Kernels}

\topar{sV×sV} \cref{fig:res_cc_svXsv_speedup} shows the speedup of our 16-bit \textsc{sssr} \verb|sV×sV| kernel over its \textsc{base} counterpart for vector densities of $0.03$-\SI{30}{\percent} and vector length 60\si{\kilo\relax}. We observe speedups of $3.0$-7.7\x, increasing near-symmetrically with the density of both operands. Speedups are generally higher for similar densities; as densities diverge, speedups converge to 5.0\x. 

To understand these trends, we first consider two steady-state edge cases. While encountering \emph{no} intersections and scanning one vector's nonzeros, \texttt{base} needs \emph{five} and \glspl{sssr} \emph{one} cycle per nonzero, explaining the 5.0\x~speedup limit for divergent densities. While processing \emph{only} intersections, \textsc{base} needs 18 and \glspl{sssr} 1.25 cycles per nonzero pair, yielding a peak achievable speedup of 14.4\x~and FPU utilization of \SI{80}{\percent} if nonzero positions were to coincide exactly. Thus, as index matching likelihood increases with both densities, so do speedups. Finally, as both vectors become very sparse and contain few elements, intersection quickly terminates and speedups are offset by \gls{sssr} setup costs. They are also modulated by the random vectors' nonzero positions and sequences, introducing slight asymmetries.

\topar{sV+sV} \cref{fig:res_cc_svPsv_speedup} shows the speedup of our 16-bit \textsc{sssr} \verb|sV+sV| kernel over \textsc{base} for the same vectors as for \verb|sV×sV|. Speedups range from 5.4 to 9.8\x~and increase with the density of both operands with slight operand asymmetry.

To understand these trends, we again consider edge cases. While reading only summand $a$ or $b$, \glspl{sssr} process each element up to 9.6\x~and 8.8\x~faster assuming continuous index comparison. In practice, occasional \gls{essr} index write backpressure decreases speedups to the observed 9.0\x and 8.2\x. Speedups are asymmetric \revone{in $a$ and $b$} as ternary branching code in \textsc{base} incurs either one or two branch cycles. For continuous index matches, peak speedups are again 14.4\x.  Thus, speedups increase with increasing density in both operands and for similar densities.

\topar{sM×sV}  \cref{fig:res_cc_smXsv_speedup} shows the speedup of our 16-bit \textsc{sssr} \texttt{sM×sV} kernel over \textsc{base}  against the average nonzeros per matrix row $\overline{n}_{nz}$ for different vector densities. We again assume that the \gls{tcdm} stores the full matrix. 

We observe similar trends as for the \texttt{sV×dV} kernel we iterate: speedups increase with the density of both arguments and approach those of \texttt{sV×dV} for corresponding densities, reaching up to 6.3\x. However, the smaller dense vector dimensions result in an earlier onset of offsetting and modulating effects as densities decrease: for \SI{0.1}{\percent} vector density or $2$-$3$ nonzeros, speedups are significantly reduced and strongly vary with nonzero patterns. However, speedups remain above 1, meaning we still amortize \gls{sssr} setup even for few nonzeros in both arguments.

\subsection{Cluster Performance}
\label{sec:results_cl}

To evaluate \revone{multi-core and} system-level performance, we implement parallel scaleouts of \verb|sM×dV| and \verb|sM×dV| on a Snitch cluster. We reuse our \revone{architecture-optimized} single-core kernels, \revone{distribute} dynamically sized chunks of rows among cores, and \revone{implement} a double-buffered data movement scheme for the matrix using the cluster \gls{dma}. 

\revone{To ensure realistic memory throughput, latency, and scheduling, we connect our cluster to one of eight channels of a Micron MT54A16G808A00AC-36 HBM2E \gls{dram} (\SI{3.6}{\giga b\per\second\per pin} or \SI{57.6}{\giga\byte\per\second} peak bandwidth, \SI{88}{\nano\second} average round-trip latency\footnote{\revone{Averaged from \gls{dma} matrix chunk reads in \gls{sssr} \texttt{sM×dV} using peak-speedup matrix \texttt{mycielskian12}. Excludes on-chip interconnect.}}), which we simulate using \emph{DRAMSys}~\cite{Jung2015DRAMSysAF}.} All input \revone{data and instructions} initially reside in \revone{this \gls{dram}} and \revone{all} results are written back to it. \revone{A \SI{16}{\kibi\byte} 4-way L2 instruction cache, bypassed by \gls{dma} data reads and writes, caches instruction fetches.} \revone{In addition to DRAM, PHY, and controller delays modeled by \emph{DRAMSys}, we model the delay of an on-chip interconnect with 16 additional cycles of forward and backward latency. We further explore the impact of interconnect latency and DRAM channel bandwidth sharing in \Cref{sec:results_cl_bwlat}.}

\topar{sM×dV} \cref{fig:res_cl_csrmv_speedup} shows the speedups of the 16-bit \textsc{sssr} kernel over the \textsc{base} kernel on \verb|sM×dV| parallelized on a cluster. Improvements are notable even for $\overline{n}_{nz}=1$ at \revone{1.7\x}~and reach up to \revone{4.9}\x, sustaining over 4\x~for $\overline{n}_{nz}>30$. 
We observe the same trends as for single-core \verb|sM×dV|, but with reduced speedups and stronger variations, which we attribute to multiple causes.
Most importantly, \emph{bank conflicts}, aggravated by the pseudorandom access patterns of indirection, lower the achievable \gls{tcdm} bandwidth and thus \gls{issr} throughput. Additionally, the \emph{initial vector transfer} cannot be overlapped with useful computation, modulating speedups with the vector's length. Our matrix \emph{double-buffering} and \emph{row distribution} schemes also incur some computational imbalance and overhead. %
\revone{The high latency, limited throughput, and scheduling of our \emph{\gls{dram} model}, though mostly hidden through the double-buffered \gls{dma} transfer of long, contiguous matrix chunks, also slightly decrease speedups by a median of \SI{1.8}{\percent} and up to \SI{6.9}{\percent}.} Finally, we observe occasional stalls due to instruction cache misses. %
Despite these inherent parallelization \revone{and scaleout} overheads, \glspl{sssr} enable up to \revone{\SI{46.8}{\percent}} overall \gls{fpu} utilization and extreme efficiency gains at minimal area cost: \emph{eight} cores with \glspl{sssr} achieve the same computational throughput as \revone{\emph{39}} cores running \textsc{base}.

\topar{sM×sV} \cref{fig:res_cl_csrmspv_speedup} shows 16-bit \textsc{sssr} \verb|sM×sV| speedups for different vector densities, which peak at \revonemod{5.9\x}. As with \verb|sM×dV|, we see similar trends as in the used single-core kernel with diminished speedups due to initial vector copy, instruction cache and \gls{tcdm} stalls, \revone{work-sharing imbalances, and \gls{dram} inefficiencies.}
\revone{For high vector densities, performance is mainly affected by \gls{tcdm} stalls and setup overheads:} the peak speedup is only \revone{\SI{6.9}{\percent}} lower than in the single-core case. 
Because each core processes only a subset of matrix rows \revone{and 
due to variations in \gls{dram} response time}, few-nonzero modulation is more pronounced overall.
\revone{As the vector density $d_v$ decreases and computational imbalance increases, \gls{dram} inefficiencies increase: for $d_v=\SI{30}{\percent}$, speedups are only \SI{0.40}{\percent} lower on average than with an ideal memory system, while for $d_v=\SI{0.1}{\percent}$, they are \SI{16}{\percent} lower}.
The minimum observed speedup of \revone{1.1\x}~is \emph{higher} than in the single-core case, which we attribute to the notably smaller code size and thus fewer instruction cache stalls the \textsc{sssr} kernel incurs.

\subsubsection{\revone{Bandwidth and Latency Sensitivity}}
\label{sec:results_cl_bwlat}

\begin{revonefig}[t]%
  \vspace{0.4em}%
  \subfloat[Cluster \texttt{sM×$\ast$V} speedups vs. limited DRAM channel bandwidth. The red dots indicate where speedups have degraded by \SI{10}{\percent}.]{%
    \hspace{0.002\linewidth}
    \includegraphics[width=0.975\linewidth]{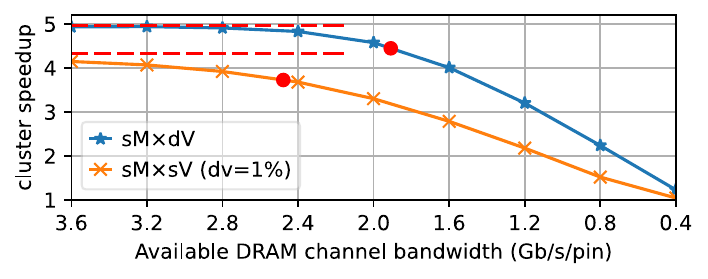}%
    \label{fig:res_cl_csrmv_bw}%
  }\\
  \subfloat[Cluster \texttt{sM×$\ast$V} speedups vs. on-chip interconnect latency.]{%
    \hspace{0.002\linewidth}
    \includegraphics[width=0.975\linewidth]{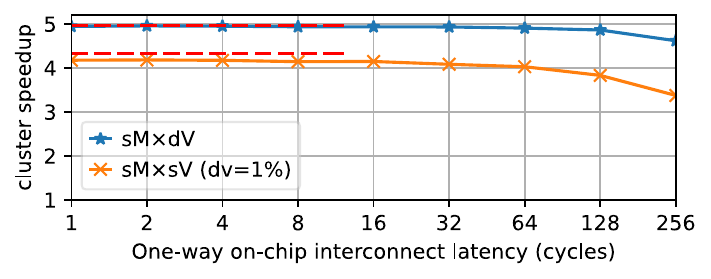}%
    \label{fig:res_cl_csrmv_lat}%
  }
  \caption{Sensitivity of \gls{sssr} Snitch cluster speedups to limited DRAM channel bandwidth and varying on-chip interconnect latency. The red dashed lines show speedups for unlimited bandwidth and zero latency.}
  \label{fig:res_cl_bwlat}
\end{revonefig}

\revone{To evaluate the sensitivity of \gls{sssr} speedups to bandwidth sharing and interconnect scaling, we rerun cluster \texttt{sM×dV} and \texttt{sM×sV} with limited \gls{dram} throughput and varying interconnect latencies. We use the peak-\texttt{sM×dV}-speedup matrix \texttt{mycielskian12} ($\overline{n}_{nz} = 133$, \SI{4.3}{\percent} density) to ensure high \gls{dram} pressure and use \SI{1}{\percent} vector density for \texttt{sM×sV}. In both cases, we also indicate ideal speedups assuming a memory system with unlimited bandwidth and zero latency.}

\revone{\Cref{fig:res_cl_csrmv_bw} shows how speedups decrease as the available \gls{dram} channel bandwidth is restricted from its full throughput of $3.6$ to \SI{0.4}{\giga b\per\second\per pin}, simulating channel sharing with other bus agents. We maintain the one-way interconnect latency of 16 cycles. %
For \gls{sssr}-accelerated \texttt{sM×dV} with full \gls{dram} bandwidth, the cluster incurs an average throughput of $R_T = \SI{1.6}{\giga b\per\second\per pin}$; we thus expect our accelerated \texttt{sM×dV} to be compute-bound above $R_T$ and memory-bound below it. %
Indeed, speedups start decreasing linearly below $R_T$ until they reach 1\x~just below \SI{0.4}{\giga b\per\second\per pin}, past which both the accelerated and baseline variants are memory-bound and perform the same. %
However, we also observe some speedup degradation \emph{above} $R_T$; this is because \gls{dram} bandwidths higher than $R_T$ can hide minor throughput variations due to computational imbalances and \gls{dram} scheduling. %
For \texttt{sM×sV}, we observe similar trends: in this case, $R_T$ is higher due to faster processing of the matrix fiber, and speedups in the compute-bound region degrade slightly faster due to increased computational imbalance%
.}

\revone{\Cref{fig:res_cl_csrmv_lat} shows how speedups decrease as on-chip interconnect latency increases, assuming full \gls{dram} channel throughput. For \texttt{sM×dV}, speedups remain near-constant until one-way latency exceeds 64 cycles, comparable to that of the downstream \gls{dram} subsystem. This is because the cluster's \gls{dma} engine moves matrix data in double-buffered chunks tens of \si{\kibi\byte} in length, which saturate the bus for hundreds of cycles; performance is negatively impacted only once the total access latency becomes comparable to the durations of these transfers. %
For \texttt{sM×sV}, the same trends hold as matrix data is still moved in large, double-buffered chunks. However, greater variations in computation rate and imbalance result in slightly more pronounced speedup losses as latency increases.}

\revone{%
Overall, the efficiency of \glspl{sssr} in removing instruction bottlenecks moves sparse workloads  %
closer to the memory bound, and they still provide notable speedups well into the memory-bound regime and only fully lose their benefits when  \gls{dram} throughput is reduced to an order of magnitude less than regular channel bandwidth. %
This is despite \texttt{sM×dV} and \texttt{sM×sV} being very memory-intensive; sparse workloads allowing for matrix data reuse, such as matrix-matrix multiplications and iterative solvers, are significantly less memory-bound. %
Furthermore, our double-buffered data movement approach proves highly latency-resilient, compensating for hundreds of cycles in interconnect latency in addition to \gls{dram} delays with minimal speedup losses.
}

\subsection{Area and Timing}
\label{sec:results_synth}

\begin{figure}[t]
  \subfloat[Area breakdown of the default streamer and major subcomponents (areas in \si{\kilo\GE}). The residual area is shown rightmost in light gray.]{
    \label{fig:res_synth_at}
    \includegraphics[width=\linewidth]{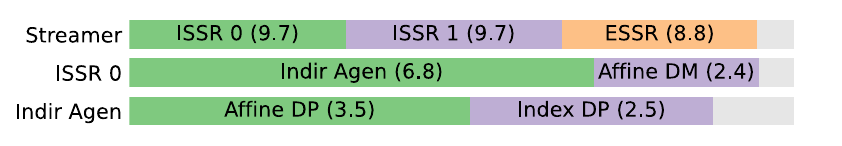}
  }\hfill
  \subfloat[Area and min. clock period for different streamer configurations.]{
    \label{fig:res_synth_atcmp}
    \includegraphics[height=3.8cm]{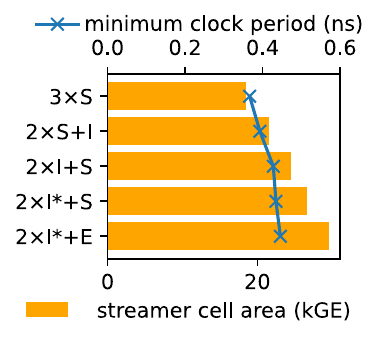}
  }
  \subfloat[Streamer area vs. clock period for different target clocks.]{
    \label{fig:res_synth_areahier}
    \includegraphics[height=3.8cm]{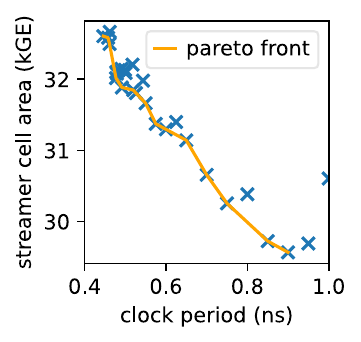}
  }\hfill
  \caption{GF12LP+ synthesis area and timing results for \gls{sssr} streamer. S: \gls{ssr}, I: \gls{issr}, E: \gls{essr}, I*: \gls{issr} with shared comparator.}
  \label{fig:res_synth}
\end{figure}

We synthesize the \gls{sssr} streamer in various configurations using Synopsys \emph{Design Compiler} for GlobalFoundries’ 12LP+ FinFET technology. We target the typical (TT) process corner at \SI{25}{\celsius} with \SI{0.8}{\volt} supply voltage. Unless stated otherwise, a \SI{1}{\giga\hertz} clock and IO delays of \SI{35}{\percent} and \SI{60}{\percent} on the core and \gls{tcdm} interfaces are constrained. In our default streamer configuration, each of the two  \glspl{issr} and single \gls{essr} has four data FIFO stages, 17-bit addresses to cover our \SI{128}{\kibi\byte} \gls{tcdm}, and four affine loop levels.

\cref{fig:res_synth_areahier} shows the default streamer's hierarchical area distribution. Each \gls{issr} contributes \SI{9.7}{\kGE} and the \gls{essr} \SI{8.8}{\kGE} to the \SI{30}{\kGE} streamer. \gls{issr} area is dominated by its address generator, only \SI{37}{\percent} of which is fully dedicated to index streaming. \cref{fig:res_synth_atcmp} compares the area and minimum achievable clock period for different streamer configurations. If only indirection capabilities are needed, only \SI{3.0}{\kGE} (\SI{16}{\percent}) in additional area is incurred per \gls{issr}. Intersection between two SSRs incurs another \SI{2.1}{\kGE}. The full \gls{sssr} streamer with added union capability incurs \SI{11}{\kGE} (\SI{60}{\percent}) area overhead over the baseline and only moderately increases the minimum clock period from \SI{367}{\pico\second} to \SI{446}{\pico\second}, still easily meeting Snitch's \SI{1}{\giga\hertz} clock target. \cref{fig:res_synth_at} shows the achieved minimum period and area for different period goals, demonstrating that \gls{sssr} streamer area scales gracefully as timing pressure increases. At the cluster level, adding \gls{sssr} streamers to all eight worker cors incurs only a minimal \SI{1.8}{\percent} cell area overhead compared to providing only regular \glspl{ssr}.

\subsection{Energy and Power}
\label{sec:results_power}

We estimate the power and total energy consumption of a Snitch cluster running our workloads from \cref{sec:results_cl} with 16-bit-index \textsc{base} and \textsc{sssr} kernels. \revone{We use the same interconnect and main memory models as in \cref{sec:results_cl}}. We target the TT process corner of  GlobalFoundries’ 12LP+ FinFET technology at \SI{1}{\giga\hertz}.
We implement the cluster using Synopsys \emph{Fusion Compiler} and estimate power for the low- and high-efficiency matrices \verb|cryg2500| and \verb|cavity12| using Synopsys \emph{PrimeTime}, then scale dynamic power with component utilizations measured in RTL simulation.

\cref{fig:res_pow_csrmv} shows the total energies for \verb|sM×dV| using both kernel variants for each matrix. While the median cluster power is predictably lower for the \textsc{base} kernel \revone{(\SI{195}{\milli\watt}} vs. \revonemod{\SI{285}{\milli\watt}}), \glspl{sssr} reduce the minimum energy per \texttt{fmadd} from \revonemod{\SI{282}{\pico\joule}} to \revone{\SI{103}{\pico\joule}} and achieve energy efficiency improvements of up to \revonemod{2.9\x}.
\cref{fig:res_pow_csrmspv} shows the total energies for \texttt{sM×sV} for a vector density of \SI{1}{\percent}. Again, \glspl{sssr} incur higher median power, but lower the minimum energy per matrix nonzero from \revonemod{\SI{107}{\pico\joule}} to \revone{\SI{43}{\pico\joule}} and achieve energy efficiency improvements of up to \revonemod{3.0\x}.

\begin{revonefig}[t]%
  \vspace{-0.6em}%
  \hspace{0.75ex}%
  \subfloat[\texttt{sM×dV}]{%
    \centering%
    \includegraphics[width=0.48\linewidth]{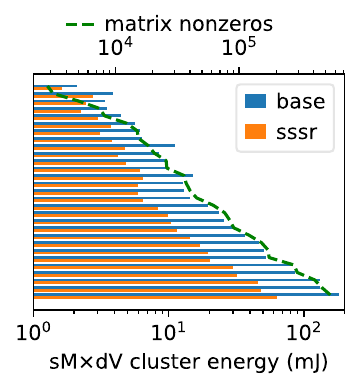}%
    \label{fig:res_pow_csrmv}%
  }\hfill%
  \subfloat[\texttt{sM×sV} (\SI{1}{\percent} vector density)]{%
    \centering%
    \includegraphics[width=0.48\linewidth]{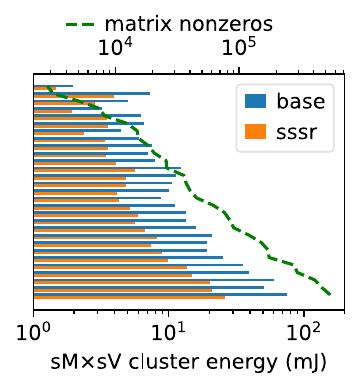}%
    \label{fig:res_pow_csrmspv}%
  }\hspace{0.75ex}
  \caption{Snitch Cluster workload energy estimates in GF12LP+. The benchmarks and data are the same as for the results in \Cref{sec:results_cl}.}
  \label{fig:res_cl_pow}
\end{revonefig}

\section{Related Work}
\label{sec:relwork}

\revone{In this section, we compare our approach to numerous state-of-the-art CPU, GPU, and accelerator solutions, covering both hardware and software proposals. We summarize our findings in two tables. \cref{tbl:relwork_comp_sw} collects the evaluation platform, matrix format, and highest fraction of peak compute achieved on all discussed FP64 \texttt{sM×dV} software implementations. \cref{tbl:relwork_comp} compares the features, flexibility, and architectural cost of all discussed hardware designs.}

\revone{\topar{Scalar processor architectures} %
A processor with \glspl{sssr} can be considered an extended and specialized \gls{dae} architecture \cite{Smith1984DecoupledAC}. 
\gls{dae} architectures  combine two decoupled processors, one handling \emph{memory accesses} and another \emph{execution}, which communicate through register-mapped \glspl{fifo}. 
\glspl{sssr} %
provide a lightweight access processor design with multiple performance enhancements.
They leverage multi-element \emph{streams} as abstractions, which %
improves %
decoupling, minimizes %
offloads from the main instruction stream, and attains near-ideal compute and memory throughputs.
Their job types define a compact, orthogonal instruction set that minimizes access code complexity and enables lightweight implementation.
Unlike the original \gls{dae} concept, \glspl{sssr} support %
multiple concurrent streams on different registers.
Our low-cost sparsity extensions maintain high throughput on indirect and joint streams without the hardware and bookkeeping overheads of a more conventional access processor architecture, enabling efficient general sparse compute without major execute processor modifications.
}

\newcommand{\rot}[1]{\adjustbox{angle=-30,llap,lap=1em}{#1}\hspace{1.5em}}
\newcommand{\tm}{\checkmark}
\newcommand{\lH}{H}
\newcommand{\lM}{M}
\newcommand{\lL}{L}

\revone{\topar{Vector processor architectures}} %
Indirection in \glspl{issr} is similar to \emph{scatter-gather} in vector architectures, which finds recent adoption in superscalar out-of-order processors. Arm's Scalable Vector Extensions \cite{stephens2017sve} and Intel's \gls{knl} \cite{sodani2016knl} add support for \gls{simd} scatter-gather. However, the short lengths and alignment requirements of \gls{simd} vectors limit scatter-gather benefits on sparse workloads compared to \glspl{issr}.
The \gls{uve} \cite{Domingos2021UnlimitedVE} relaxes vector length constraints and introduces vector-register-bound indirect streams similar to \glspl{issr}. However, it requires vectorizable workloads and cannot operate efficiently at scalar granularity.
Gong et al. \cite{Gong2020SAVESV} extend a vector multiply-add unit to skip ineffectual and coalesce useful computations. While this is orthogonal to \glspl{sssr}, it is only effective on low, regular sparsity in uncompressed data.

\revone{\topar{Software optimizations for scalar and vector processors} Numerous software solutions aim to accelerate sparse \gls{la} on scalar and vector processors.
The CVR matrix format \cite{Xie2018CVREV} aims to improve \texttt{sM×dV} \gls{simd} lane and cache efficiency on x86 CPUs. Still, it utilizes at most \SI{0.69}{\percent} of peak FP64 compute on Xeon Phi 7250.
Zhang et al.~\cite{Zhang2018VectorizedPS} optimize \texttt{sM×dV} %
for Intel's AVX-512 extensions, proposing a format based on \gls{sell} that enables peak FP64 compute utilizations of up to \SI{1.5}{\percent} on Xeon Phi 7230.
Regu2D~\cite{Fei2021Regu2DAV} leverages an adaptive 2D partitioning scheme and arranges rows as similar-sized groups with zero-padding as necessary to fill \gls{simd} lanes, improving their platform's peak FP64 utilization to \SI{3.1}{\percent}.
Alappat et al.~\cite{Alappat2020PerformanceMO} model the performance of \texttt{sM×dV} on Fujitsu's A64FX using the \gls{sell}-C-\textsigma~matrix format and report up to \SI{130.9}{\Gflops} or \SI{4.7}{\percent} of peak FP64 compute. Nevertheless, this is 9.9\x~less than what we achieve on a Snitch cluster with \glspl{sssr}.}
\mbox{Z. Xie} et al. \cite{Xie2022APS} present a \revone{\gls{csr}-interfaced} \verb|sM×sM| library for multi- and manycores selecting the best internal format and algorithm with a deep learning model based on nonzero patterns.

\begin{table}
\setlength{\tabcolsep}{0.6em}
\centering
\ifx\showrebuttal\undefined
\else
\color{blue}
\fi
\begin{tabular}{cp{7.5em}llr}
\toprule
 & Work &%
 Platform &%
 Format &%
 Peak FP Ut. \\
 \midrule
 \multirow{4}{*}{\adjustbox{angle=90}{CPU}}%
 & CVR~\cite{Xie2018CVREV}                        & Xeon Phi 7250  & CVR               & \SI{0.69}{\percent} \\
 & Zhang et al.~\cite{Zhang2018VectorizedPS}      & Xeon Phi 7230  & SELL-like         & \SI{1.5}{\percent}  \\
 & Regu2D~\cite{Fei2021Regu2DAV}                  & Xeon Gold 6132 & Regu2D            & \SI{3.1}{\percent}  \\
 & Alappat et al.~\cite{Alappat2020PerformanceMO} & A64FX          & SELL-C-\textsigma & \SI{4.7}{\percent}  \\
 \midrule
 \multirow{6}{*}{\adjustbox{angle=90}{GPU}}%
 & Tsai et al.~\cite{Tsai2020SparseLA}            & V100           & CSR             & \SI{1.6}{\percent}  \\
 & Merrill et al.~\cite{Merrill2016MergeBasedPS}  & K40            & CSR             & \SI{2.0}{\percent}  \\
 & TileSpMV~\cite{Niu2021TileSpMVAT}              & A100           & tile-adapt.     & \SI{2.9}{\percent}  \\
 & Tsai et al.~\cite{Tsai2020SparseLA}            & Radeon VII     & CSR             & \SI{3.2}{\percent}  \\
 & cuSPARSE~\cite{nvidia2018cuda}                 & GTX 1080 Ti    & CSR             & \SI{17}{\percent}   \\
 & TileSpMV~\cite{Niu2021TileSpMVAT}              & Titan RTX      & tile-adapt.     & \SI{27}{\percent}   \\
 \midrule
 & \textbf{SSSRs (ours)}                        & Snitch + \glspl{sssr} & CSR & \SI{47}{\percent} \\
\bottomrule
\end{tabular}
\caption{\revone{Overview of discussed FP64 \texttt{sM×dV} software implementations.}}
\label{tbl:relwork_comp_sw}
\end{table}

\topar{Streaming \gls{isa} extensions} %
General-purpose processor extensions increasingly leverage streams as abstractions to accelerate compute.
Prodigy \cite{Talati2021ProdigyIT} uses compile-time-inferred metadata to prefetch nested indirect streams. However, it does not eliminate load-store or bookkeeping instructions.
SpZip \cite{Yang2021SpZipAS} reads and writes compressed irregular data using decoupled streaming units. This is similar to \glspl{sssr} reading and writing fiber-based formats, but SpZip does not support intersection or union, %
and a fetcher-compressor pair incurs \SI{116}{\kGE} or 3.9\x~more area than an \gls{sssr} streamer.
Z. Wang et al. \cite{Wang2019StreambasedMA} propose a decoupled register-mapped extension accelerating affine and indirect memory streams similar to \glspl{issr}. However, streams need to stepped explicitly, severely limiting possible speedups on in-order cores, and intersection and union are not supported.
SparseCore \cite{Rao2022SparseCoreSI} accelerates intersection, union, and subtraction of indexed streams and provides dedicated functional units. However, indirection is not accelerated, making the handling of one-sided sparsity inherently inefficient. Additionally, SparseCore incurs \SI{0.183}{\meter\meter^2}  per stream unit in Open-Cell \SI{15}{\nano\meter}, which is comparable to two entire Snitch \glspl{cc} with \glspl{sssr}.

\begin{table}
\setlength{\tabcolsep}{0.6em}
\centering
\begin{tabular}{cp{11.5em}cccccr}
\toprule
 & \vspace{2.3em}\adjustbox{lap=1em}{Work} &%
 \rot{\,~~Open source} &%
 \rot{\,~1-sided accel.} &%
 \rot{\,~2-sided accel.} &%
 \rot{~~Usage flexib.\textsuperscript{*}} &%
 \rot{Sparsity flexib.\textsuperscript{\textdagger}} &%
 \rot{~~\si{\kGE} per core\textsuperscript{\textdaggerdbl}} \\
 \midrule
 \multirow{8}{*}{\adjustbox{angle=90}{CPU}}%
 & SVE S/G \cite{stephens2017sve}               &     & \tm &     & \lM & \lH & -  \\
 & KNL S/G \cite{sodani2016knl}                 &     & \tm &     & \lM & \lH & -  \\
 & UVE \cite{Domingos2021UnlimitedVE}           & \tm\textsuperscript{**} & \tm &     & \lH & \lH & 72\textsuperscript{\P} \\
 & Gong et al. \cite{Gong2020SAVESV}            &     & \tm & \tm & \lM & \lL & 31\textsuperscript{\P} \\
 & Prodigy \cite{Talati2021ProdigyIT}           &     & \tm &     & \lM & \lH & 10\textsuperscript{\P} \\
 & SpZip \cite{Yang2021SpZipAS}                 &     & \tm &     & \lM & \lH & 116 \\
 & Z. Wang et al. \cite{Wang2019StreambasedMA}  &     & \tm &     & \lH & \lH & -   \\
 & SparseCore \cite{Rao2022SparseCoreSI}        &     &     & \tm & \lH & \lH & 619 \\
 \midrule
 \multirow{3}{*}{\adjustbox{angle=90}{GPU}}%
 & A100 \cite{nvidia2020a100}                   &     & \tm &     & \lM & \lL & -   \\
 & Zhu et al. \cite{zhu2019sparsetensorcore}    &     & \tm &     & \lL & \lM & 12\textsuperscript{\S}  \\
 & Y. Wang et al. \cite{Wang2021DualsideST}     &     &     & \tm & \lL & \lM & 157\textsuperscript{\S} \\
 \midrule
 \multirow{7}{*}{\adjustbox{angle=90}{Accelerators}}%
 & MatRaptor \cite{Srivastava2020MatRaptorAS}   &     &     & \tm & \lL & \lH & - \\
 & OuterSPACE \cite{Pal2018OuterSPACEAO}        &     &     & \tm & \lL & \lH & - \\
 & Sadi et al. \cite{Sadi2019EfficientSO}       &     & \tm &     & \lL & \lH & - \\
 & EIE \cite{Han2016EIEEI}                      &     & \tm & \tm & \lL & \lH & - \\
 & ExTensor \cite{Hegde2019ExTensorAA}          &     &     & \tm & \lM & \lH & - \\
 & Dadu et al. \cite{Dadu2019TowardsGP}         &     & \tm & \tm & \lM & \lH & - \\
& \revone{SISA \cite{Besta2021SISASI}}          &     &     &\revone{\tm} & \revone{\lM} & \revone{\lH} & - \\
 \midrule
 & \textbf{SSSRs (ours)}                        & \tm & \tm & \tm & \lH & \lH & 30 \\
 
\bottomrule
\end{tabular}
\\[0.5em]
\justifying\noindent
L/M/H = Low/Medium/High. *\,Qualitative, based on operand granularity, use constraints, overheads. \textdagger\,Qualitative, based on efficiently handled range, structure. \textdaggerdbl\,Based on authors' data, different cores used. **\,gem5 model only, no RTL. \P\,Arch. state only at 6T=1.5GE per bit. \S\,Only node given; we assume GlobalFoundries' 22nm and 12nm, resp.
\caption{Overview of discussed hardware \revone{designs}.}
\label{tbl:relwork_comp}
\end{table}

\topar{GPU software} %
\revone{Moving from general-purpose processors to more specialized architectures, GPUs have been the subject of intensive investigations aimed at improving their efficiency on sparse data computations.}
One-sided sparsity on GPUs is often accelerated in \emph{software} through specialized algorithms and tensor formats.
Merrill et al. \cite{Merrill2016MergeBasedPS} improve the performance consistency of \gls{csr} \verb|sM×dV| on GPUs using a merge-based decomposition, \revone{reaching up to \SI{2.0}{\percent} of peak FP64 compute on a K40 GPU}.
Nvidia provides optimized sparse \gls{la} kernels for their GPUs through \emph{cuSPARSE} \cite{nvidia2018cuda}. We evaluate their \gls{csr} \verb|sM×dV| kernels from CUDA Toolkit 10.0 on a Jetson AGX Xavier (FP32 only) and a GTX 1080 Ti GPU (FP32 and FP64), profiling 100 invocations with \emph{nvprof} on the same matrices as for \glspl{sssr}. For FP32, both GPUs exhibit high peak \gls{sm} occupancies of \SI{96}{\percent} and \SI{87}{\percent}, but low peak floating-point utilizations of and \SI{2.1}{\percent} and \SI{0.75}{\percent} on active \glspl{sm}, suggesting low thread parallelism among warps. FP64 on the GTX 1080 Ti shows similar \gls{sm} occupancies, but a notably higher peak floating-point utilization of \SI{17}{\percent}, likely due to the 32\x~fewer FP64 cores per \gls{sm}. %
\revone{Tsai et al. \cite{Tsai2020SparseLA} create optimized \texttt{sM×dV} library kernels for both Nvidia and AMD GPUs and various formats. Their \gls{csr} kernels regularly outperform vendor code and reach up to \SI{1.6}{\percent} and \SI{3.2}{\percent} of peak FP64 utilization on Nvidia V100 and AMD Radeon VII GPUs, respectively.
TileSpMV \cite{Niu2021TileSpMVAT} uses warp-optimized kernels on sparse tiles in a given or adaptively selected format. It achieves peak FP64 utilizations of \SI{2.9}{\percent} on A100 and \SI{27}{\percent} on Titan RTX, which again has 32\x~fewer FP64 than FP32 cores per \gls{sm}.
Still, this is 1.7\x~lower than in a Snitch cluster with \glspl{sssr}.
}
Shi et al. \cite{Shi2020EfficientSM} accelerate \verb|sM×dM| by encoding matrices as arrays of coordinate-encoded tiles, \revone{improving} access contiguity and data reuse. However, the average FP32 speedups of 1.7\x~over cuSPARSE on SuiteSparse matrices are insufficient to reach FP utilizations comparable to those of Snitch with \glspl{sssr}.
Karimi et al. \cite{Karimi2022VCSRAE} \revone{design} a set of memory-aware matrix formats accelerating \verb|sM×dV| by improving cache usage and thread-level parallelism. While improving performance, their best format achieves median FP32 speedups of only 1.8\x~(K40) and 3.5\x~(V100) over cuSPARSE \gls{csr} on Suite- Sparse matrices, still not enough to match our utilizations.

Some sparse-sparse software approaches for GPUs also exist: Zachariadis et al. \cite{Zachariadis2020AcceleratingSM} accelerate \verb|sM×sV| using a tensor-core-based outer-product approach with dense 16\x16 tiles, accepting low flexibility and losses of unnecessary computations within blocks.
Li et al. \cite{Li2020AdaptiveSO} optimize general sparse matrix-vector multiply by dynamically choosing one of eight \verb|sM×dV|/\verb|sM×sV| approaches based on tensor properties using a \gls{ml} model. This solution inherently cannot exceed the peak performance of the existing approaches chosen from, but can be adapted to improve overall performance on other systems including ones with \glspl{sssr}.

\topar{GPU hardware} %
Extensions to GPU \emph{hardware} accelerating sparsity have also been proposed. 
Nvidia's A100 architecture \cite{nvidia2020a100} introduces support for one-sided \emph{structured} sparsity by efficiently handling up to two zeros in every block of four values. \glspl{sssr} efficiently handle a much wider range of sparsities ($\gg\;$\SI{50}{\percent}) without any structural requirements and enable random accesses into a full \gls{tcdm}.
Zhu et al. \cite{zhu2019sparsetensorcore} present an algorithm and co-optimized tensor core modifications for efficient sparse-dense neural network inference.
Y. Wang et al. \cite{Wang2021DualsideST} extend V100 tensor cores to efficiently handle sparse-sparse matrix multiply and convolution with an outer product approach. However, the bitmap encoding used hinders efficient performance and memory scaling beyond low sparsities, and the dense output allows for the overall approach to be replicated with scattering \glspl{issr}.

\topar{Hardware accelerators} Sparse accelerators usually target a single application domain like \gls{dl} \cite{Dave2021HardwareAO} or specific \gls{la} operations and dataflows.
Most use custom data formats and precisions tailored to their use case and are incapable of general-purpose computation. Notably, a general-purpose processor with \glspl{sssr} can implement many of the employed specialized computation schemes.
MatRaptor \cite{Srivastava2020MatRaptorAS} and OuterSPACE \cite{Pal2018OuterSPACEAO} accelerate \verb|sM×sM| with row-wise and outer product dataflows, respectively; \glspl{sssr} can efficiently handle both dataflows.
Sadi et al. \cite{Sadi2019EfficientSO} accelerate large-scale \verb|sM×dV| by column-slicing matrices and then accumulating the sparse results, improving vector locality and reuse; iterating over slices could be done on a host core while \glspl{sssr} handle multiplication and result accumulation.
Han et al.'s EIE \cite{Han2016EIEEI} minimize DRAM access energy in \gls{dl} inference by keeping full, weight-shared models in on-chip SRAM; \glspl{sssr} also enable high utilization on sparse data with codebooked values, but are not limited to one application, schedule, or dataflow.
ExTensor \cite{Hegde2019ExTensorAA} hierarchically intersects \gls{csf} tensors to avoid redundant computation and transfers; a processor with \glspl{sssr} can replicate this by handling the upper axes in software while the streamer intersects the performance-critical primary axes.
Dadu et al. \cite{Dadu2019TowardsGP} present a compute fabric accelerating both stream joins and streaming indirection. However, the loosely-coupled spatial compute approach incurs notable costs in area and programmability.
\revone{SISA \cite{Besta2021SISASI} introduces a set-centric \gls{isa} to accelerate graph mining algorithms. In addition to set intersection and union, it also accelerates set difference, membership and size queries, and element insertion or removal. It supports both indexed and bitmap sets and a binary-search-based "galloping" mode for index joins between sets of divergent sizes. While SISA is highly specialized for graph mining and does not target one-sided sparsity or "low-complexity" algorithms, many of its additional capabilities could be added to SSSRs. However, SISA's architectural cost is discussed only in loose analogy to other designs, making such tradeoffs hard to assess.}

\section{Conclusion}
\label{sec:conclusion}

In this work, we extend \glspl{ssr} to accelerate \emph{indirection}, \emph{intersection}, and \emph{union}, creating a modular, backward-compatible \gls{isa} extension enabling efficient general sparse linear algebra on sparse-fiber-based formats including the widespread \gls{csr} and \gls{csf}. 
We evaluate our sparse \glspl{ssr} in the RISC-V Snitch system by presenting efficient single-core sparse-dense and sparse-sparse \gls{la} kernels and parallelizing matrix-vector products on an eight-core cluster.

In a single core using 16-bit indices, the streaming indirection of \glspl{sssr} enables peak \verb|sV×dV| \gls{fpu} utilizations of \SI{80}{\percent} and \verb|sM×dV| speedups of up to 7.0\x~over our optimized baseline. 
The intersection and union of \glspl{sssr} enable speedups of $3.0$-$7.7$\x~and $5.4$-$9.8$\x~on sparse-sparse dot product and vector addition across a wide sparsity range, enabling \verb|sM×sV| speedups of up to 6.3\x. 
Scaled out to a Snitch cluster \revone{with interconnect and DRAM models exhibiting realistic latency, throughput, and scheduling}, \verb|sM×dV| using \glspl{sssr} is up to \revone{4.9\x}~faster and \revonemod{2.9\x}~more energy efficient, while \verb|sM×sV| is up to \revonemod{5.9\x}~faster overall and \revonemod{3.0\x}~more energy efficient for a vector density of \SI{1}{\percent}.
These improvements incur only \SI{11}{\kGE} in additional area per core or \SI{1.8}{\percent} across a cluster, which may further be reduced thanks to the streamer's modular design.

Unlike recent CPU, GPU, and accelerator proposals, \glspl{sssr} accelerate one- \emph{and} two-sided sparse compute while remaining flexible in dataflow and operand sparsity and incurring a minimal area impact. A Snitch cluster with \glspl{sssr} running \verb|sM×dV| achieves \revone{9.9\x}~and \revonemod{1.7\x}~higher peak floating-point utilizations than recent CPU and GPU software.


\section*{Acknowledgement}

\revone{We thank the reviewers for their valuable feedback.}
This work was supported by the ETH Future \mbox{Computing} Laboratory (EFCL), financed by a donation from Huawei Technologies.
It also received funding from the European High-Performance Computing Joint Undertaking (JU) under grant agreement No 101034126 (The European Pilot) and Specific Grant Agreement No 101036168 (EPI SGA2).

\bibliographystyle{IEEEtran}
\bibliography{IEEEabrv,main}
\begin{IEEEbiography}[\includegraphics{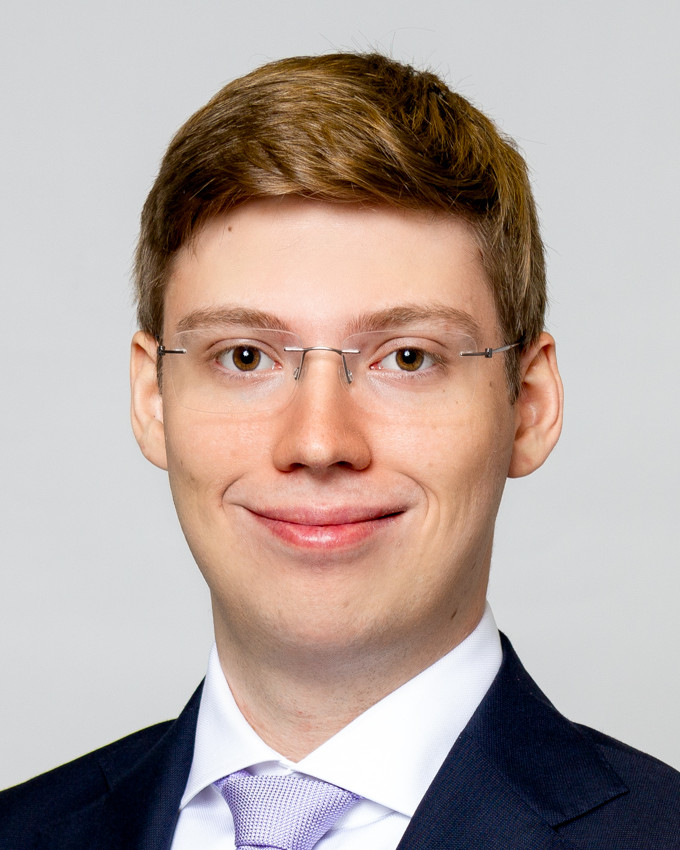}]{Paul Scheffler}
received his BSc and MSc degree in electrical engineering and information technology from ETH Zurich in 2018 and 2020, respectively. He is currently pursuing a PhD in the Digital Circuits and Systems group of Prof. Benini. His research interests include hardware acceleration of sparse and irregular workloads, on-chip interconnects, manycore architectures, and high-performance computing. 
\end{IEEEbiography}
\begin{IEEEbiography}[\includegraphics{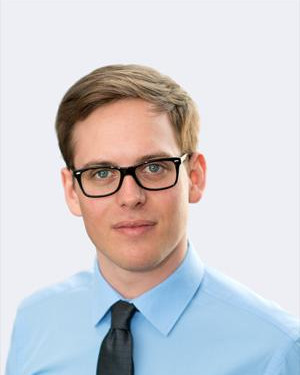}]{Florian Zaruba}
received his BSc degree from TU Wien in 2014, his MSc and PhD from the Swiss Federal Institute of Technology Zurich in 2017 and 2021. He is a system architect at Axelera AI.  His research interests include design of very large scale integration circuits and high performance computer architectures.
\end{IEEEbiography}
\begin{IEEEbiography}[{\includegraphics[width=1in]{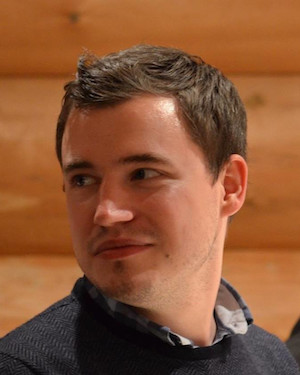}}]{Fabian Schuiki}
received the BSc, MSc, and PhD degree in electrical engineering from the Swiss Federal Institute of Technology Zürich (ETHZ), in 2014, 2016, and 2021, respectively. He is a hardware compiler engineer at SiFive Inc. His research interests include transprecision computing as well as near- and in-memory processing.
\end{IEEEbiography}
\vfill
\newpage
\begin{IEEEbiography}[\includegraphics{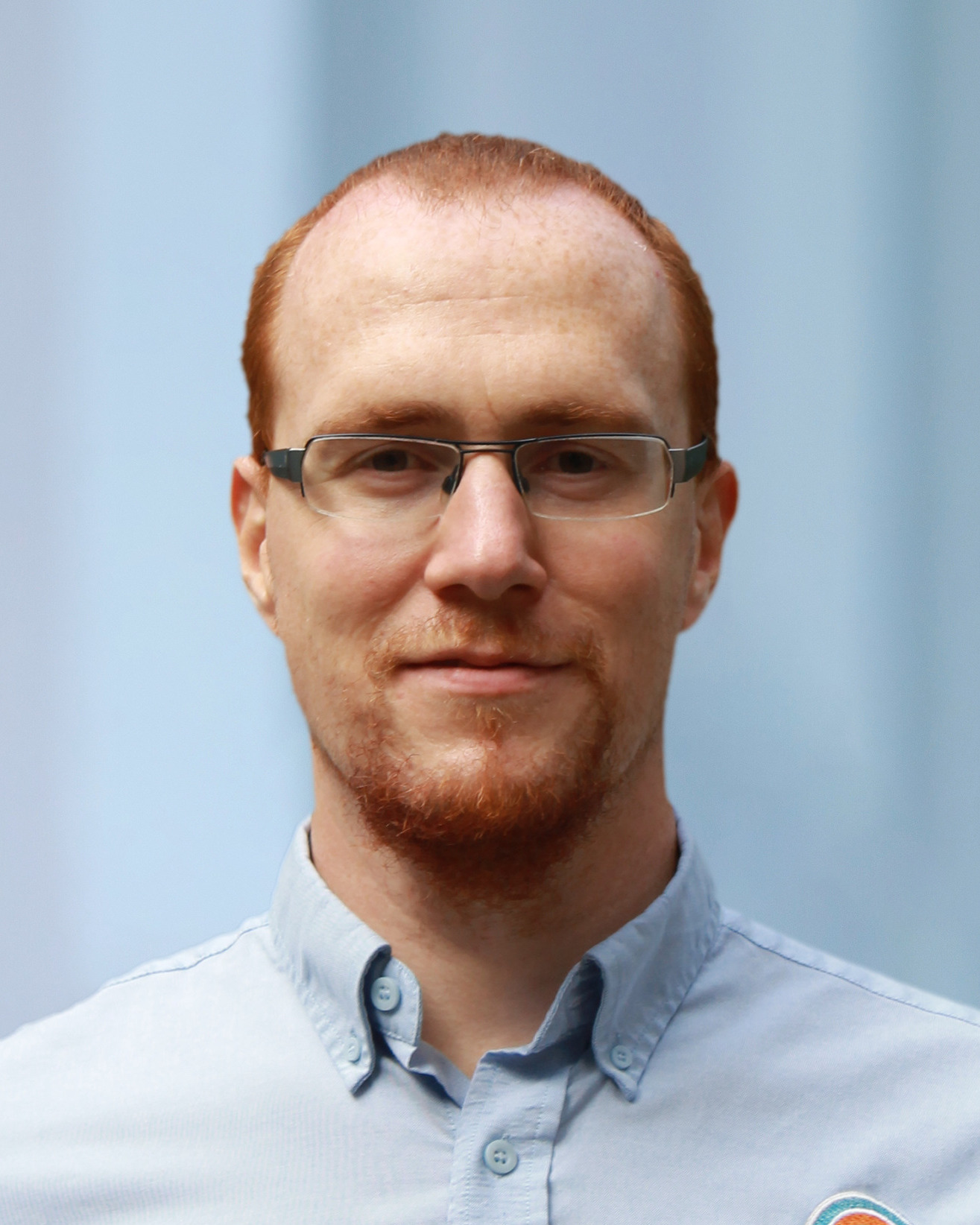}]{Torsten Hoefler}
is a Professor of Computer Science at ETH Zürich, Switzerland. He received his PhD from Indiana University. Dr. Hoefler's interests are around performance-centric software and hardware development. He is a Fellow of the ACM and a member of the Academia Europaea.
\end{IEEEbiography}
\begin{IEEEbiography}[\includegraphics{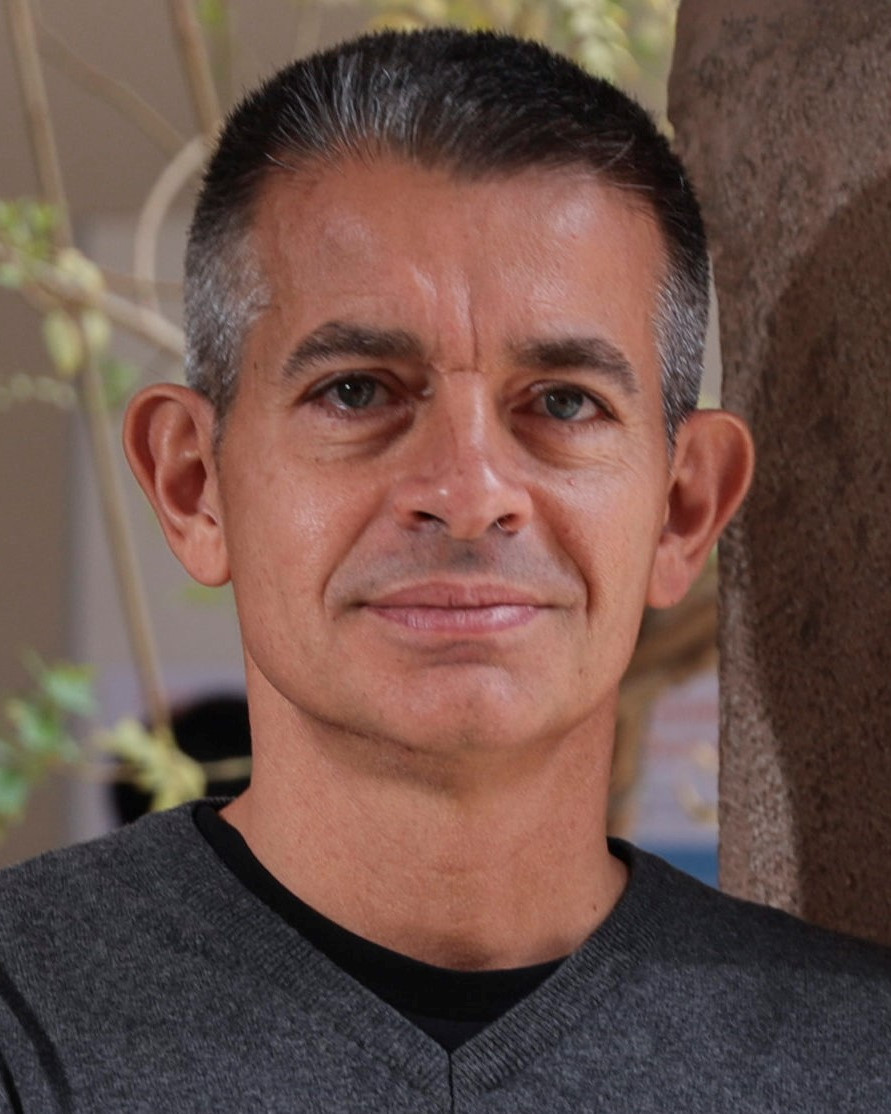}]{Luca Benini}
holds the chair of digital Circuits and systems at ETHZ and is Full Professor at the Universita di Bologna. He received a PhD from Stanford University.  Dr. Benini’s research interests are in energy-efficient parallel computing systems, smart sensing micro-systems and machine learning hardware. He is a Fellow of the ACM and a member of the Academia Europaea. 
\end{IEEEbiography}
\vfill
\end{document}

%% file: acr.tex
\newacronym{hpc}{HPC}{high performance computing}
\newacronym{isa}{ISA}{instruction set architecture}
\newacronym{fpu}{FPU}{floating-point unit}
\newacronym{fp}{FP}{floating-point}
\newacronym{mac}{MAC}{multiply-accumulate}
\newacronym{muldiv}{MULDIV}{integer multiply-divide unit}
\newacronym{alu}{ALU}{arithmetic logic unit}
\newacronym{fu}{FU}{functional unit}
\newacronym{fifo}{FIFO}{first in, first out}
\newacronym{cgra}{CGRA}{coarse-grained reconfigurable array}
\newacronym{pim}{PIM}{processing-in-memory}
\newacronym{dae}{DAE}{decoupled access/execute}
\newacronym{dram}{DRAM}{dynamic random access memory}
\newacronym{sell}{SELL}{sliced ELLPACK}

\newacronym[firstplural=Core Complexes (CCs)]{cc}{CC}{core complex}
\newacronym{dmcc}{DMCC}{data movement CC}
\newacronym{fpss}{FPSS}{floating-point subsystem}
\newacronym{ssr}{SSR}{stream semantic register}
\newacronym{issr}{ISSR}{indirection SSR}
\newacronym{essr}{ESSR}{egress SSR}
\newacronym{sssr}{SSSR}{sparse stream semantic register}
\newacronym{tcdm}{TCDM}{tightly-coupled data memory}
\newacronym{frep}{FREP}{floating-point repetition}
\newacronym{dma}{DMA}{direct memory access}

\newacronym{csf}{CSF}{compressed sparse fiber}
\newacronym{csr}{CSR}{compressed sparse rows}
\newacronym{csc}{CSC}{compressed sparse columns}
\newacronym{dcsr}{DCSR}{doubly-compressed sparse rows}
\newacronym{bcsr}{BCSR}{blocked compressed sparse rows}
\newacronym{sdcsr}{sDCSR}{sliced doubly-compressed sparse rows}
\newacronym{spmv}{SpMV}{sparse matrix-vector multiplication}
\newacronym{sdmv}{SDMV}{sDCSR matrix-vector multiplication}

\newacronym{spvv}{SpVV}{sparse vector-vector multiplication}
\newacronym{csrmv}{CsrMV}{CSR matrix-vector multiplication}
\newacronym{csrmm}{CsrMM}{CSR matrix-matrix multiplication}
\newacronym{cvr}{CVR}{compressed vectorization-oriented sparse row}
\newacronym{dl}{DL}{deep learning}
\newacronym{la}{LA}{linear algebra}

\newacronym{ml}{ML}{machine learning}
\newacronym{simd}{SIMD}{single-instruction, multiple-data}
\newacronym{sm}{SM}{streaming multiprocessor}
\newacronym{rtl}{RTL}{register transfer level}
\newacronym{knl}{KNL}{Knights Landing}
\newacronym{uve}{UVE}{Unlimited Vector Extension}
\newacronym{fem}{FEM}{finite element analysis}
\newacronym{loess}{LOESS}{locally estimated scatterplot smoothing}